\documentclass[]{jfm_arxiv}
\usepackage{graphicx}
\usepackage{newtxtext}
\usepackage{comment}
\usepackage{newtxmath}
\usepackage{natbib}
\usepackage{setspace}
\usepackage{subcaption}
\usepackage{hyperref}
\usepackage{tikz}
\usepackage{amsmath}
\usepackage{amssymb}
\usepackage{multirow}
\usepackage{multicol}
\usepackage[mathscr]{eucal}
\usepackage{bm}
\usepackage{eqlist} 
\usepackage{graphicx}
\usepackage{xfrac}
\usepackage{subcaption}
\hypersetup{
    colorlinks = true,
    urlcolor   = blue,
    citecolor  = blue,
}

\newcommand{\RomanNumeralCaps}[1]
\linenumbers
\usepackage{float}
\usepackage{placeins}

\usepackage{algorithm}
\usepackage{algorithmic}
\usepackage{makecell}
\usepackage{booktabs}
\usepackage{cancel}
\title{Intermittent Vortex Merging and Extreme Drag in Transitional Airfoil Flow}

\author{Shishir Gautam\aff{1}
  , \and Chitrarth Prasad \aff{1} \corresp{\email{c.prasad@okstate.edu}}}
\affiliation{\aff{1} School of Mechanical and Aerospace Engineering, Oklahoma State University, OK 74078}

\begin{document}

\maketitle

\begin{abstract}
Intermittent departures from nominal Kelvin--Helmholtz shedding can produce rare and pronounced drag excursions in transitional airfoil flow. 
We examine these events using two-dimensional direct numerical
simulations of flow over a NACA0012 airfoil at an angle of attack of $5^\circ$, a free-stream Mach number of $0.4$, and chord-based Reynolds numbers of $5\times10^4$ and $5\times10^5$. 
At the lower Reynolds number, event-resolved analysis shows that individual primary vortices are released from the separated shear layer through the eruption of wall-generated, opposite-signed secondary vorticity. 
Each eruption interrupts the connection between a developing primary vortex and its feeding shear layer, releasing the vortex downstream. 
During nominal shedding, the vortex reaching the trailing-edge region is associated with a single such release and remains sufficiently isolated to pass the trailing edge without strong collective interaction. 
Extreme events instead arise through clustered vortex release, in which several secondary-vorticity eruptions occur within a short interval and produce a compact group of primary vortices with small initial streamwise spacing.
Differential convection further reduces their spacing and promotes strong near-trailing-edge interactions, where the combined pressure footprint of these vortices produces a localized suction peak and a sharp increase in drag. 
These interactions range from prolonged deformation and filamentation to rapid core coalescence. 
Similar compact vortex organization and near-trailing-edge interactions are recovered at $Re=5\times10^5$, indicating that the downstream event pathway persists despite the smaller vortical scales.
These findings suggest that controlling vortex-release timing through
secondary-vorticity dynamics may provide a route to disrupt clustered
release and mitigate extreme aerodynamic loading.



\end{abstract}


\section{Introduction}\label{sec:Intro}
Extreme events ($EEs$) are rare, short-lived departures from typical system behavior that can have disproportionately large consequences~\citep{Extr_Events_Dynamics_Statistics_Prediction, Extreme_events_review, Route_to_EEs_Mishra,annurev_statistics_of_EE_Sapsis}.
Such events arise across natural, biological, socioeconomic, and engineering systems, including disease outbreaks~\citep{disease_outbreak1_cohen2000changing, disease_outbreak2_al2012outbreak}, financial markets~\citep{stockmarket2_vandewalle1998crash, stockmarket1_sornette2003critical}, seismic activity~\citep{seismic_activities_geller1997earthquake}, oceanic rogue waves~\citep{Rouguewaves2_kharif2003physical, Rougewaves1_dysthe2008oceanic}, and climate variability~\citep{Extreme_weather3_ropelewski1987global, Extreme_weather4_dakos2008slowing}.
Fluid systems present a particular challenge because $EEs$ often emerge from transient interactions among multiple flow structures.
Examples include intermittent acoustic bursts in turbulent jets~\citep{Schmidt_Schmid_CPOD_2019,cavalieri2010intermittent,cavalieri2011_jittering,kearney2013_intermittent_jitter}, turbulent spots in transitional boundary layers~\citep{goparaju_cpod_transition,marxen2019turbulence,simoni2016wavelet}, extreme inlet distortions in aircraft engines~\citep{tanguy2018characteristics,gil2018assessment}, sudden cavitation events~\citep{wu2021cavitation,escaler2006cavitation,ram2020mechanisms,bhatt2020numerical}, and supersonic-inlet buzz~\citep{chima2012analysis,johnson2026investigations,ye2024buzz}.
Their intermittent occurrence and short duration make it difficult to isolate the mechanisms that initiate them from the background unsteadiness of the flow.


Transitional airfoil flows are particularly susceptible to intermittent
changes in aerodynamic loading because of the unsteady dynamics of
laminar separation bubbles (LSBs). 
An LSB forms when a laminar boundary layer separates and subsequently reattaches downstream on the surface, enclosing a region
of recirculating flow. 
Intermittent bursting of the bubble can leave the downstream flow fully separated, produce low-frequency variations in the aerodynamic forces, and cause a rapid loss of lift
\citep{Sandham_2008, almutairi2010,toppings2023_transient_LSB_Exp_Fluids,
toppings_yarusvyech_2024_LSB_Burst_JFM,
toppings_yarusvych_2024_LSB_Burst}. 
The onset and persistence of such
bursting have been ascribed to several parameters including the angle of attack ($\alpha$), Reynolds number ($Re$), external forcing, and incoming disturbances, among others~\citep{gaster,marxen2011effect, SERNA201543, Michelis_Yarusevych_Kotsonis_2017, Dellacasagrande_Lengani_Simoni_Yarusevych_2023}.



Large force excursions have also been reported when the separation bubble remains nominally attached. 
For incompressible transitional flows at $Re\sim\mathcal{O}(10^4)$, \citet{Rudy} and \citet{Barthel} observed recurrent extreme excursions in the lift and drag coefficients, $C_L$ and $C_D$, and linked them to an interaction between a low-frequency component associated with the extreme events, $f_e$, and the higher-frequency vortex-shedding dynamics, $f_v$. Irregular or weakly periodic force fluctuations have likewise been reported for NACA0012 flows by \citet{jones2008_thesis} and \citet{Chiu_Tseng_Chang_Chou_2023_FPM}.

A less understood source of intermittent aerodynamic loading is variability within the Kelvin--Helmholtz (KH) driven roll-up of the separated shear layer itself. In canonical free shear layers, KH instability produces coherent vortices whose spacing, strength, and subharmonic interactions govern subsequent pairing and merging~\citep{Winant_Browand_1974, moser1993three,cavalieri2010intermittent}. 
The evolution of these structures need not remain uniform from one shedding cycle to the next. 
In subsonic jets, for example, intermittent variations in the amplitude, position, and spatial extent of KH-type wavepackets are commonly described as \textit{jitter} and have been linked to intermittent acoustic radiation~\citep{cavalieri2011_jittering, kearney2013_intermittent_jitter, towne2017_jitter}. 
These observations raise the possibility that analogous departures from nominal KH roll-up in an airfoil separation bubble may produce unusually large aerodynamic-load excursions. 
However, the event-level sequence connecting such departures to individual extreme force events remains poorly understood.


The present study investigates the mechanism by which intermittent departures from nominal KH-driven vortex shedding produce extreme drag events in transitional flow over a NACA0012 airfoil. A highly resolved two-dimensional direct numerical simulation is performed at free-stream Mach number ($M$) of $0.4$, $Re=50{,}000$, and $\alpha$ of $5^\circ$. At these conditions, previous numerical studies have reported an unsteady laminar separation bubble, KH-driven shear-layer roll-up, and irregular aerodynamic-force fluctuations~\citep{SHAN20051096, jones2008_thesis, jones2008, almutairi2010, JONES_SANDBERG_SANDHAM_2010}. The configuration therefore provides a well-documented setting to isolate the event-level dynamics responsible for the largest drag excursions. The computational methodology, grid assessment, and validation of the numerical database are presented in $\S~\ref{sec: DNS}$.

Because $EEs$ involve short-lived and spatially localized changes in the separated shear layer, the analysis follows the temporal evolution of individual events rather than relying solely on time-averaged descriptions. 
In $\S~\ref{sec:Mechanism_of_Extreme_Drag_Events}$, event-resolved flow fields, surface-pressure distributions, and vortex tracking are used to identify how intermittent departures from the nominal shedding sequence produce extreme drag excursions. 
The persistence of the identified pathway at a higher $Re$ and its implications for three-dimensional flow are then examined in $\S~\ref{sec:High_Re}$. 
The principal physical findings and their broader significance are summarized in $\S~\ref{sec: Conclusions}$.

\section{NACA0012 airfoil database} 
\label{sec: DNS}
\subsection{Governing Equations \& Test Conditions}
All simulations are performed using a parallel in-house solver for the
two-dimensional compressible Navier--Stokes equations. The equations are
written in generalized curvilinear coordinates $(\xi,\eta)$ in conservative
form as
\begin{equation}
    \frac{\partial}{\partial t}
    \left(
        \frac{\boldsymbol{Q}}{|J|}
    \right)
    +
    \frac{\partial \boldsymbol{F}}{\partial \xi}
    +
    \frac{\partial \boldsymbol{G}}{\partial \eta}
    =
    \frac{\partial \boldsymbol{F}^{v}}{\partial \xi}
    +
    \frac{\partial \boldsymbol{G}^{v}}{\partial \eta},
    \label{eq:governing_equations}
\end{equation}
where
\begin{equation}
    \boldsymbol{Q}
    =
    \begin{pmatrix}
        \rho \\
        \rho U \\
        \rho V \\
        \rho e
    \end{pmatrix}
\end{equation}
is the vector of conservative variables and
\begin{equation}
    J
    =
    \frac{\partial(\xi,\eta)}{\partial(x,y)}
\end{equation}
is the Jacobian of the coordinate transformation. Here, $\rho$ is the
fluid density, $U$ and $V$ are the Cartesian velocity components, and
$e$ is the total specific energy,
\begin{equation}
    e
    =
    \frac{T}{\gamma(\gamma-1)M^2}
    +
    \frac{1}{2}\left(U^2+V^2\right).
\end{equation}

The inviscid flux vector ($\boldsymbol{F,G}$) in either computational direction
$\kappa\in\{\xi,\eta\}$ is written compactly as
\begin{equation}
    \boldsymbol{F}^{(\kappa)}
    =
    \frac{1}{|J|}
    \begin{pmatrix}
        \rho \mathcal{U}_\kappa \\
        \rho U \mathcal{U}_\kappa+\kappa_x p \\
        \rho V \mathcal{U}_\kappa+\kappa_y p \\
        (\rho e+p)\mathcal{U}_\kappa
    \end{pmatrix},
    \label{eq:inviscid_fluxes}
\end{equation}
where
\begin{equation}
    \mathcal{U}_\kappa
    =
    \kappa_x U+\kappa_y V
\end{equation}
is the contravariant velocity in the $\kappa$ direction. Thus,
$\boldsymbol{F}^{(\xi)}=\boldsymbol{F}$ and
$\boldsymbol{F}^{(\eta)}=\boldsymbol{G}$ in
(\ref{eq:governing_equations}).

The corresponding viscous flux vector ($\boldsymbol{F^v,G^v}$) in compact form is
\begin{equation}
    \boldsymbol{F}^{v,(\kappa)}
    =
    \frac{1}{|J|}
    \begin{pmatrix}
        0 \\
        \kappa_x\tau_{xx}+\kappa_y\tau_{xy} \\
        \kappa_x\tau_{xy}+\kappa_y\tau_{yy} \\
        \kappa_x
        \left(
            U\tau_{xx}+V\tau_{xy}-q_x
        \right)
        +
        \kappa_y
        \left(
            U\tau_{xy}+V\tau_{yy}-q_y
        \right)
    \end{pmatrix},
    \label{eq:viscous_fluxes}
\end{equation}
where
\begin{equation}
    \tau_{ij}
    =
    \frac{\mu}{Re}
    \left(
        \frac{\partial u_i}{\partial x_j}
        +
        \frac{\partial u_j}{\partial x_i}
        -
        \frac{2}{3}
        \frac{\partial u_k}{\partial x_k}\delta_{ij}
    \right)
\end{equation}
and
\begin{equation}
    q_i
    =
    -\frac{\mu}
    {(\gamma-1)M^2Re\,Pr}
    \frac{\partial T}{\partial x_i}.
\end{equation}
where $i,j,k\in\{x,y\}$, $\delta_{ij}$ is the Kronecker delta, and the final term follows from Stokes' hypothesis.

The flow variables are non-dimensionalized using their corresponding
free-stream values, with the airfoil chord $c$ used as the reference
length. Pressure is scaled by $\rho_\infty U_\infty^2$, and time is
scaled by $c/U_\infty$. The temperature-dependent dynamic viscosity is
evaluated using Sutherland's law with $S=110.4~\mathrm{K}$, and the
fluid is treated as a calorically perfect gas. The dimensionless
parameters $Re$, $M$, and $Pr$ are defined using free-stream properties
and the airfoil chord, with $Pr=0.72$ and $\gamma=1.4$.

Two Reynolds numbers, $Re=50{,}000$ and $500{,}000$, are considered,
both at $M=0.4$ and an angle of attack of $5^\circ$. The corresponding
flow conditions, time-step sizes ($\Delta t$), sampling durations ($(tU_\infty/c)_s$), and numbers of
stored snapshots ($N_{\mathrm{snaps}}$) are summarized in Table~\ref{tab:CFD_parameters}.
Unless otherwise stated, the analysis in $\S~\ref{sec: DNS}$ and $\S~\ref{sec:Mechanism_of_Extreme_Drag_Events}$ focuses on the $Re=50{,}000$ case.
The persistence of the identified extreme-event mechanism at the higher $Re$ is examined separately in
$\S~\ref{sec:High_Re}$.

\subsection{Mesh Topology \& Resolution}
\begin{figure}
    \centering 
    \includegraphics[angle=90,trim=0cm 0cm 3cm 0cm,clip,width=\linewidth]{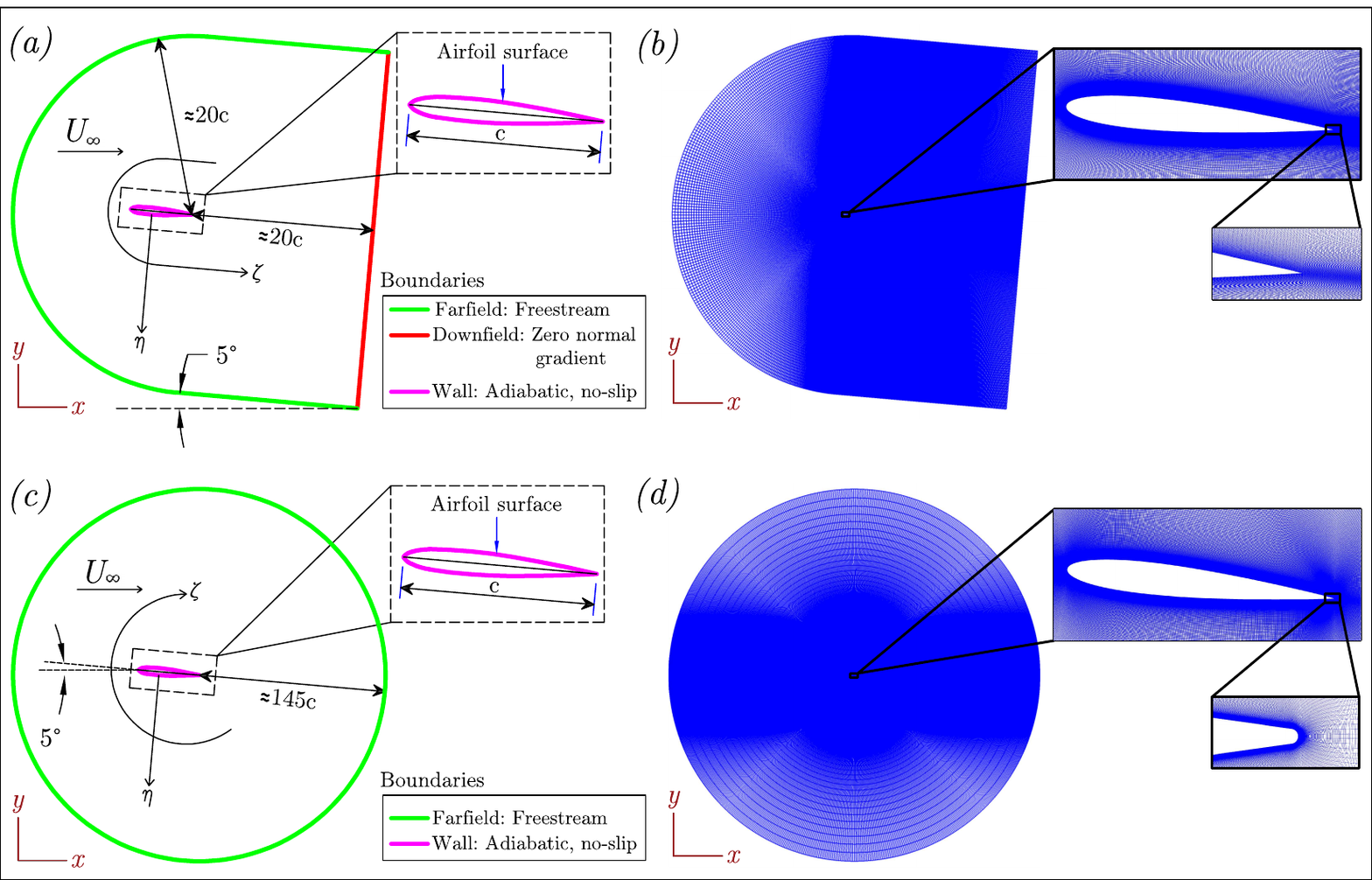}
   \caption{Computational domains and boundary conditions for the
(a,b) C-grid and (c,d) O-grid configurations. The insets show the
near-airfoil and trailing-edge mesh topology.}
    \label{fig:Domain_and_Mesh}
\end{figure}

Structured body-fitted meshes with C- and O-grid topologies are used in the present study. 
The C-grids, $G_1$ and $G_2$, represent the airfoil with a sharp trailing edge and include a wake cut extending downstream from the trailing edge. 
Grid $G_3$ is an O-grid that wraps continuously around an airfoil with a finite-thickness trailing edge. 
The computational domains for both grid topologies are shown in Fig.~\ref{fig:Domain_and_Mesh} and the metrics highlighting the grid resolutions for all three grids are summarized in Table~\ref{tab:Grid_metrics}. 

\begin{table}
    \centering
       \caption{Simulation parameters for the two test cases. The subscripts $i$ and $s$ denote the initial transient and sampling durations, respectively. } \vspace{0.2cm}
    \label{tab:CFD_parameters}
    \resizebox{\textwidth}{!}{
    \begin{tabular}{c c c c c c c c c c c } 
        Case
        & $Re$
        & $M$ 
        & $\alpha$
        & $\rho_\infty$ (kg/m$^3$) 
        & $T_\infty$ (K)
        & $c$ (m)
        & $\Delta t$ (s) 
        & $(tU_\infty/c)_i$ 
        & $(tU_\infty/c)_s$     
        & $N_\mathrm{snaps}$ \vspace{0.2cm}
        \\
        
        \textbf{$C_1$}
        & $5\times10^4$
        & 0.4
        & $5^\circ$
        & 0.18
        & 300
        & 0.036
        & $5\times10^{-7}$
        & 250
        & 722
        & 18800\\

        \textbf{$C_2$}
        & $5\times10^5$
        & 0.4
        & $5^\circ$
        & 1.84
        & 300
        & 0.036
        & $2\times10^{-7}$
        & 39
        & 138
        & 17900 \\        
    \end{tabular}
    }

\end{table}
The coarse C-grid, $G_1$,
contains $951\times251$ points, whereas the refined C-grid, $G_2$,
contains $1901\times501$ points and has twice the number of grid points above and around the airfoil surface than $G_1$. 
The O-grid, $G_3$, contains $1501\times601$ points, all of which extend circumferentially around the airfoil. 
Thus, although $G_2$ and $G_3$ have comparable total point counts, $G_3$ places substantially more points along the airfoil surface and provides much finer near-wall resolution.

\begin{figure}
    \centering 
    \includegraphics[trim=0.05cm 0.1cm 6cm 4cm,clip,width=0.5\linewidth]{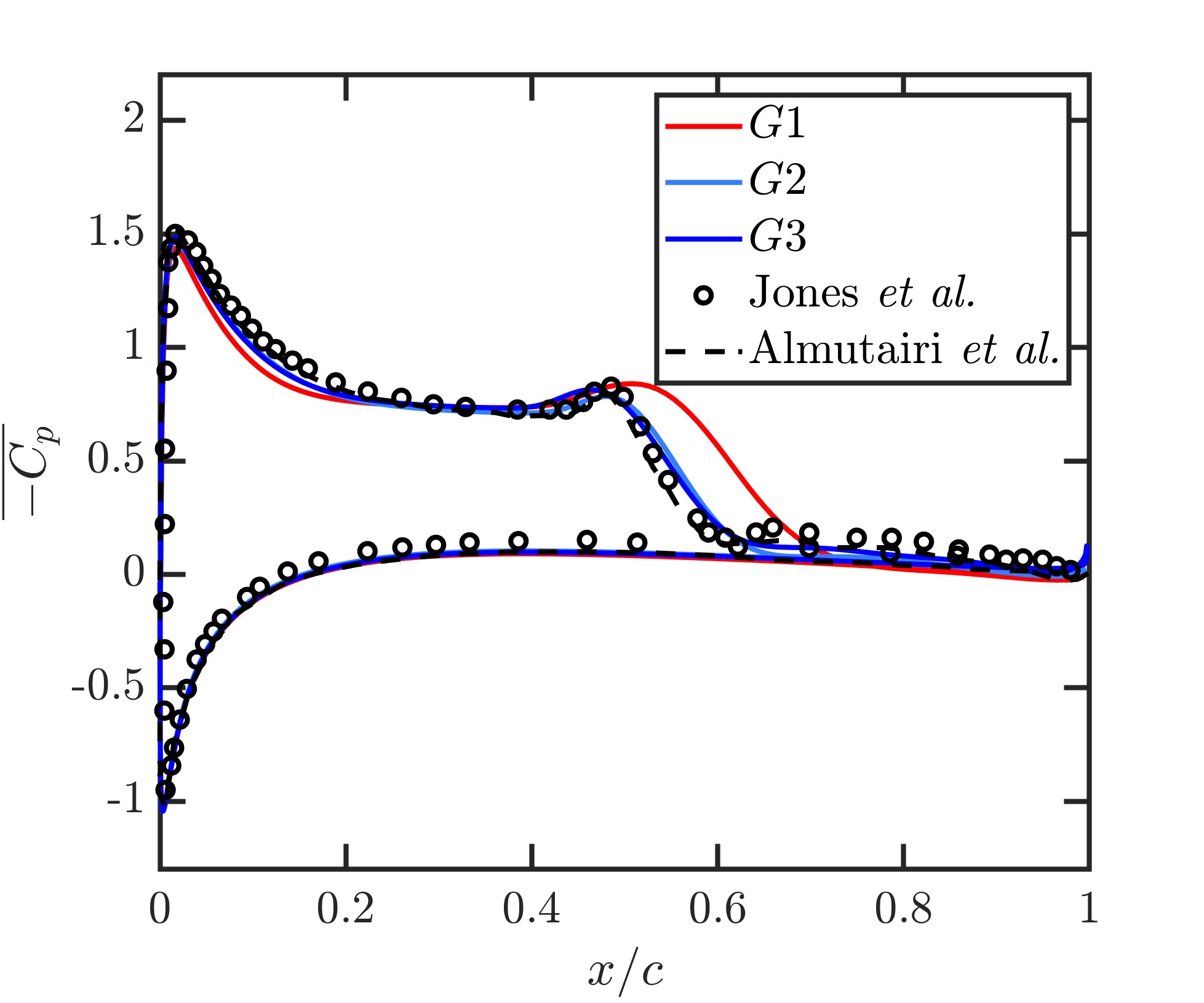}
\caption{Time-averaged surface-pressure coefficient for case $C_1$
computed on grids $G_1$--$G_3$, compared with the DNS results of
\citet{jones2008} and \citet{almutairi2010}.}
    \label{fig:grid_independency_results}
\end{figure}

Grid sensitivity is assessed by computing the $Re=50{,}000$ case 
on all three meshes. 
Because $G_1$ and $G_2$ use the same C-grid topology and sharp trailing-edge representation, their comparison provides a direct assessment of spatial refinement. 
As shown in Fig.~\ref{fig:grid_independency_results}, the time-averaged surface-pressure coefficient obtained using $G_1$
deviates noticeably from the converged results of $G_2$ and $G_3$.
Refinement to $G_2$ substantially improves the prediction, and the resulting $\overline{C_p}$ distribution agrees closely with the reference DNS results of \citet{jones2008, almutairi2010}.
Case $C_1$ is also computed on $G_3$ to assess the sensitivity of the solution to additional near-wall refinement, mesh topology, and trailing-edge representation. 
In particular, the minimum wall-normal spacing ($\mathrm{min}(\Delta s/c)_{\eta,a}$) above the airfoil surface decreases from
$1.3\times10^{-4}$ on $G_2$ to $9.8\times10^{-7}$ on $G_3$, while the number of surface points increases from 501 to 1501. 
Despite these substantial differences, the $\overline{C_p}$ distributions obtained using $G_2$
and $G_3$ are nearly indistinguishable, and both agree closely with
the reference DNS results, as shown in Fig.~\ref{fig:grid_independency_results}.

Based on these comparisons, grid $G_2$ is used for the detailed analysis of case $C_1$. 
Grid $G_3$ is used for case $C_2$ because its greater surface resolution and smaller minimum wall-normal spacing are better suited to resolving the thinner boundary layer at $Re=500{,}000$. 
As demonstrated later in $\S~\ref{sec:High_Re}$, the same extreme-event mechanism is recovered for this O-grid configuration, indicating that the mechanism is not specific to the sharp trailing-edge representation used for $C_1$.

\begin{table}
    \centering
    \caption{Comparison of grid metrics used for the numerical simulations. The subscripts $g$ and $a$ indicate the metrics evaluated on the total grid and on the airfoil surface, respectively. Refer to figure~\ref{fig:Domain_and_Mesh} for the notation of $\zeta$ and $\eta$.} \vspace{0.2cm}
    \resizebox{\textwidth}{!}{
    \begin{tabular}{c c c c c c c}
        Grid-ID      
        
        & $(N_{\zeta} \times N_{\eta} )_{\mathrm{g}}$ 
        
        & $\text{min}(\Delta s/c)_{\mathrm{{\zeta,a}}}$ 

        & $\text{max}(\Delta s/c)_{\mathrm{{\zeta,a}}}$ 
        
        & $\text{min}(\Delta s/c)_{\mathrm{\eta,a}}$ 

        & $\text{max}(\Delta s/c)_{\mathrm{\eta,a}}$ 
        &$(N_{\zeta})_{\mathrm{a}}$\vspace{0.2cm}\\

        G$_1$
        & 951 $\times$ 251 
        & 2.8 $\times$ 10$^{-4}$
        & 1.8 $\times$ 10$^{-2}$
        & 2.6 $\times$ 10$^{-4}$
        & 2.8 $\times$ 10$^{-4}$
        & 251  \\

        G$_2$
        & 1901 $\times$ 501  
        & 1.4 $\times$ 10$^{-4}$
        & 8.9 $\times$ 10$^{-3}$
        & 1.3 $\times$ 10$^{-4}$
        & 1.4 $\times$ 10$^{-4}$
        & 501  \\

        G$_3$
        & 1501 $\times$ 601  
        & 1.9 $\times$ 10$^{-4}$
        & 2.9 $\times$ 10$^{-3}$
        & 9.8 $\times$ 10$^{-7}$
        & 1.6 $\times$ 10$^{-6}$
        & 1501 \\
    \end{tabular}  
    }
    
    \label{tab:Grid_metrics}
\end{table}


\subsection{Numerical Discretization \& Boundary Conditions}

Spatial derivatives are evaluated using a fourth-order compact finite-
difference scheme~\citep{lele1992compact,gaitonde1998high}. 
A sixth-order compact Pad\'e type filter~\citep{gaitonde2000pade} is used to attenuate high-wavenumber numerical content and maintain stability of the compact discretization~\citep{visbal2002large,garmann2013comparative}.
Time advancement is performed using the second-order implicit diagonalized Beam--Warming scheme~\citep{beam1978implicit,
pulliam1981diagonal}.

The boundary layers are resolved directly without a wall model.
No-slip, adiabatic conditions are imposed at the airfoil surface.
Free-stream conditions are prescribed at the outer boundaries, while a zero-normal-gradient condition is applied at the downstream boundary.
The outer boundaries are placed sufficiently far from the airfoil to minimize their influence on the near-field flow.

The simulations are initialized from a uniform free-stream state without imposed disturbances and advanced until a self-sustained unsteady flow is established. 
The initial transient is excluded from the analysis,
and data are collected over the sampling intervals summarized previously in Table~\ref{tab:CFD_parameters}. The selected sampling frequencies and
durations provide sufficient temporal resolution to capture the short-lived flow structures associated with the extreme events.

\subsection{Unsteady Aerodynamic Loads}

\begin{figure}
    \centering 
        \begin{subfigure}[c]{\linewidth}  
            \centering 
            \includegraphics[angle=90,trim=0.1cm 0.2cm 8.5cm 0.15cm, clip, width=\linewidth]{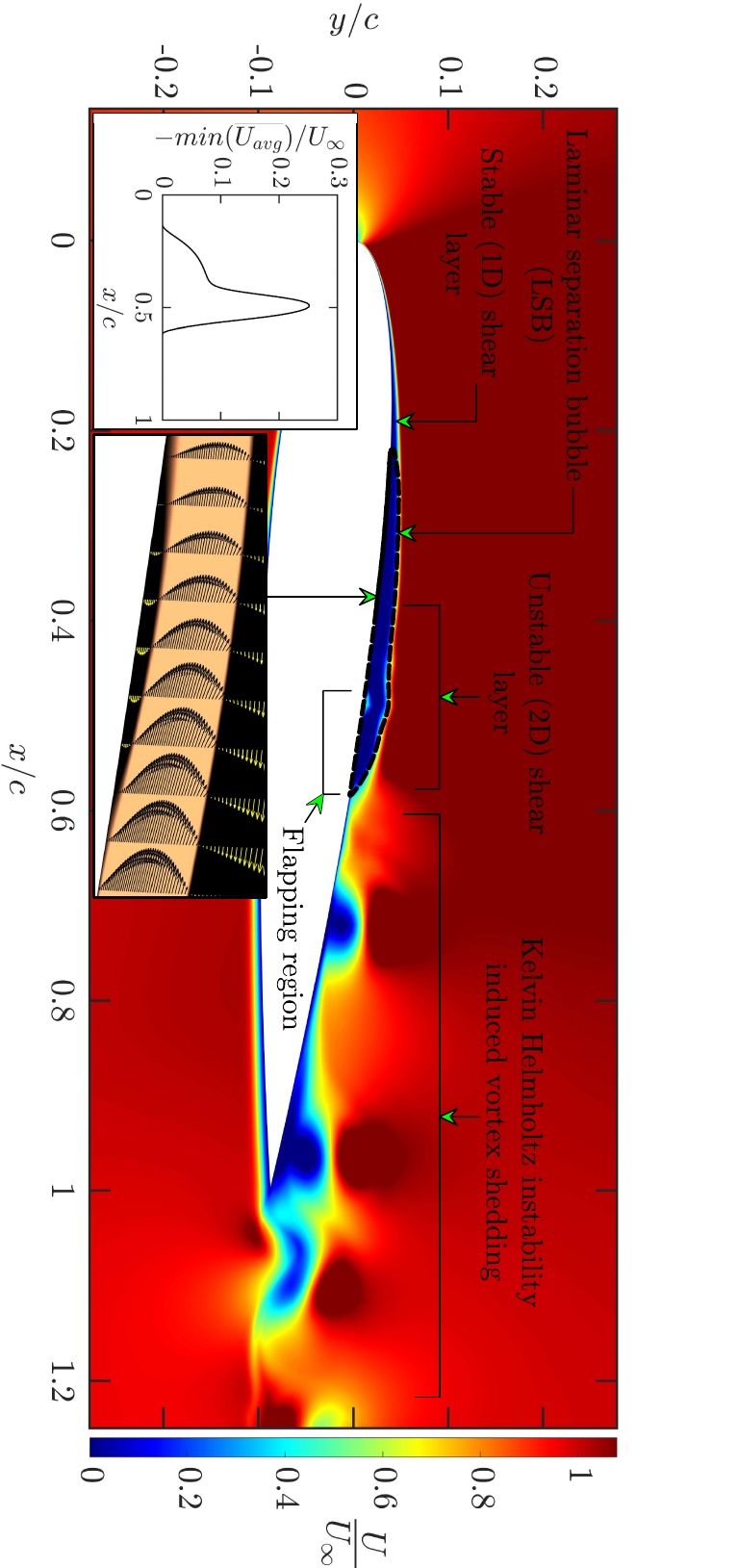}
            \label{fig: 1}            
        \end{subfigure} \vspace{-0.5cm}
\caption{Instantaneous streamwise velocity, $U/U_\infty$, for case
$C_1$ at $tU_\infty/c \approx 250$. The dashed contour identifies the
instantaneous separated-flow region. The lower-left inset shows the
time-averaged reversed-flow region on the suction surface, and the
second inset shows velocity vectors across the inflectional separated
shear layer.}
    \label{fig:flow_anatomy}    
\end{figure}

The instantaneous streamwise-velocity field in Fig.~\ref{fig:flow_anatomy} shows a separated region over the suction surface due to a strong adverse pressure gradient (APG) condition. This causes the accelerated flow near the leading edge to separate before reattaching near mid-chord, forming a closed recirculating zone or LSB.


The location and extent of the time-averaged separation bubble are identified from the region of negative streamwise velocity
\citep{ALAM_SANDHAM_2000,SHAN20051096}, as shown in the graphical inset of Fig.~\ref{fig:flow_anatomy}. 
The bubble extends over approximately one-half of the airfoil chord and has the maximum height of 0.9\% of $c$ at $x/c \approx 0.43$. Similarly, the LSB has the maximum reverse velocity of 25.2\% of freestream velocity observed at $x/c \approx 0.49$. 
The local velocity vectors shown in the other inset of Fig.~\ref{fig:flow_anatomy} depict an inflectional shear layer above the bubble, which is susceptible to convective KH instability~\citep{ALAM_SANDHAM_2000, SHAN20051096, Theofilis,rist,
He_Gioria_Perez_Theofilis_2017}. 
Roll-up of this shear layer produces the recurring vortex-shedding dynamics that define the baseline flow state from which the $EEs$ (examined later) deviate.


As discussed in $\S~\ref{sec:Intro}$, KH-driven shear layers can exhibit intermittent
departures from their otherwise recurring evolution
\citep{cavalieri2010intermittent}. 
In the present flow, this intermittency appears as variations in the formation time, spacing,
strength, and interaction of successive vortices, producing irregular changes in the aerodynamic loading.

\begin{figure}
    \centering
    \begin{tabular}[c]{@{}c@{}}   
        \begin{subfigure}[c]{\linewidth}
            \centering 
            \includegraphics[angle=90,trim=0.55cm 0.75cm 10.8cm 0.55cm, clip, width=\linewidth]{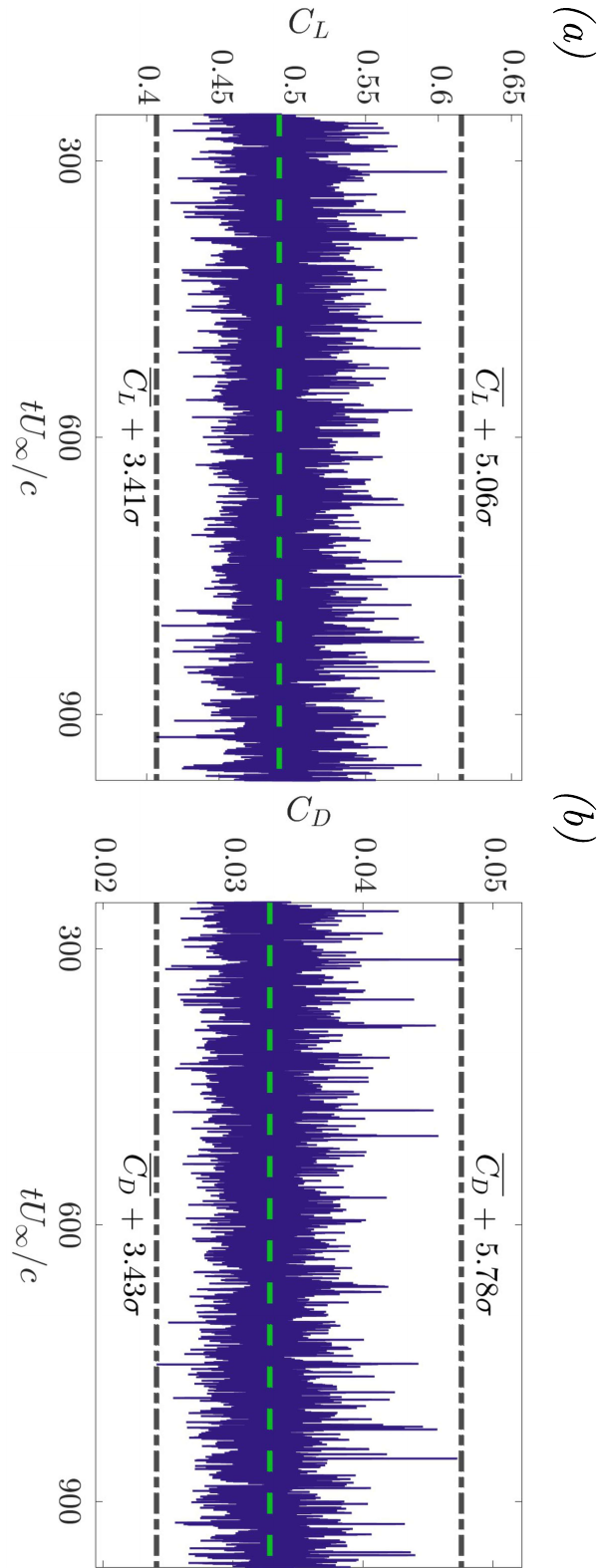}
            \label{fig: Coeff_1}
            \vspace{-0.5cm}
        \end{subfigure} 
    \end{tabular}
\caption{Time histories of (a) lift coefficient, $C_L$, and
(b) drag coefficient, $C_D$, for case $C_1$. Dashed lines denote the
mean values and dash-dotted lines denote the extrema over the sampled interval.}
    \label{fig:Aero_Coeffs_Main}    
\end{figure}

The corresponding time histories of the lift and drag coefficients are shown in Fig.~\ref{fig:Aero_Coeffs_Main}. 
Both $C_L$ and $C_D$ exhibit irregular fluctuations punctuated by intermittent excursions from their respective mean values. 
The dashed lines indicate the means, while the
dash-dotted lines mark the maximum positive and negative deviations observed over the sampled interval. 
Although pronounced excursions occur
in both types of deviation, the present study focuses on the positive drag excursions because they represent transient aerodynamic penalties.  
The accompanying lift response is retained
to characterize the overall aerodynamic signature of these events. 
The broadband frequency content underlying the irregular load fluctuations
is examined in the following subsection.

\subsection{Spectral Analysis}
To characterize the temporal scales represented in the numerical database, power spectral densities of the aerodynamic loads are estimated using Welch's method. 
The collected $N_\mathrm{snaps}$  are divided into segments of $1{,}024$ samples with $50\%$ overlap, and frequency is reported using Strouhal number
$St=fc/U_\infty$.

\begin{figure}
    \centering 
    \begin{subfigure}[c]{0.48\linewidth} 
        \includegraphics[angle = 90,trim= 0.1cm 12.5cm 1.5cm .15cm, clip, width =.9\textwidth]{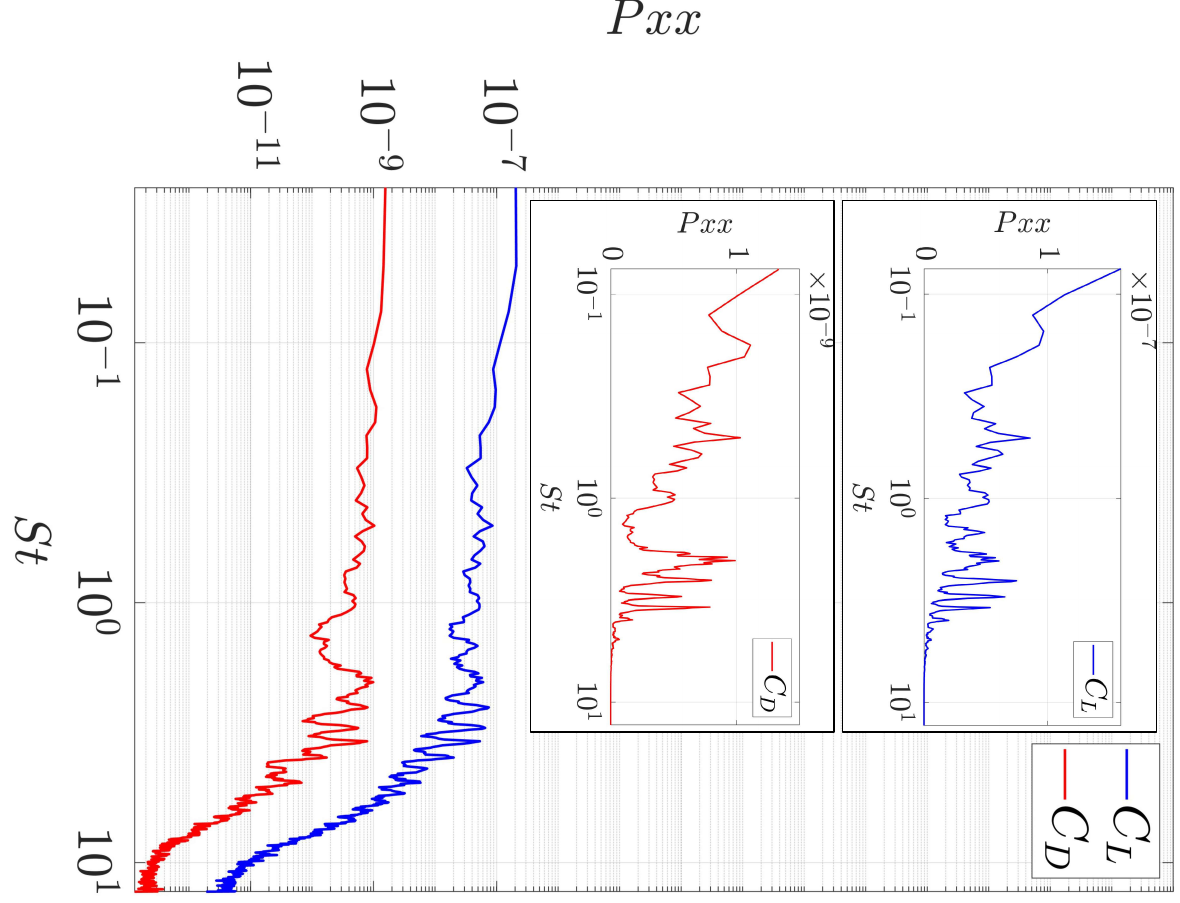}
    \end{subfigure}
\caption{Power spectral densities of the lift and drag coefficients
for case $C_1$. Insets show the same spectra on a semi-logarithmic
frequency axis.}
    \label{fig:psd}
\end{figure}

 The $C_L$ and $C_D$ spectra in Fig.~\ref{fig:psd} exhibit broadband content together with a concentration of energy over approximately
$2.5\lesssim St\lesssim4$. 
The spectral energy decreases toward the highest resolved frequencies, with no accumulation of energy near the sampling limit. 
Thus, the intermittent load excursions are embedded
within a broadband, dynamically resolved signal rather than appearing as isolated high-frequency fluctuations. 
The broad shedding-related
band is consistent with the cycle-to-cycle variability of the KH-driven vortex formation and interaction discussed in the preceding subsection.
 

\begin{figure}
    \centering   
    \begin{subfigure}[c]{1\linewidth}
        \centering
        \includegraphics[trim=8cm 17.7cm 6cm 21.3cm,clip, width =1\textwidth]{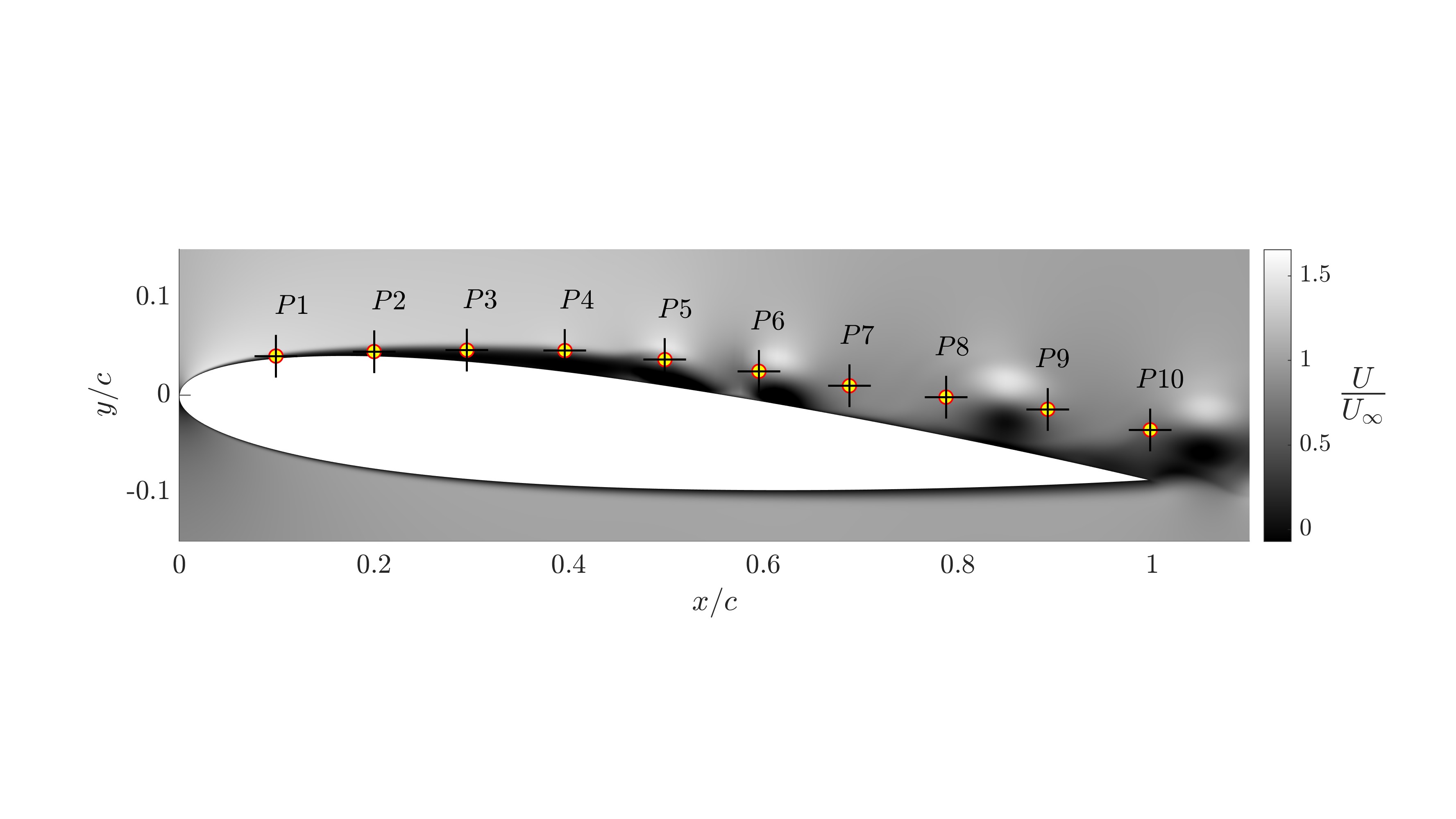}
        \subcaption[]{Probe stations}
        \label{fig:probe_location} \vspace{0.2cm}
    \end{subfigure}
    \begin{subfigure}[c]{.25\linewidth}
        \centering 
         \includegraphics[trim=3.9cm 1cm 5cm 2.65cm, clip, width =\linewidth]{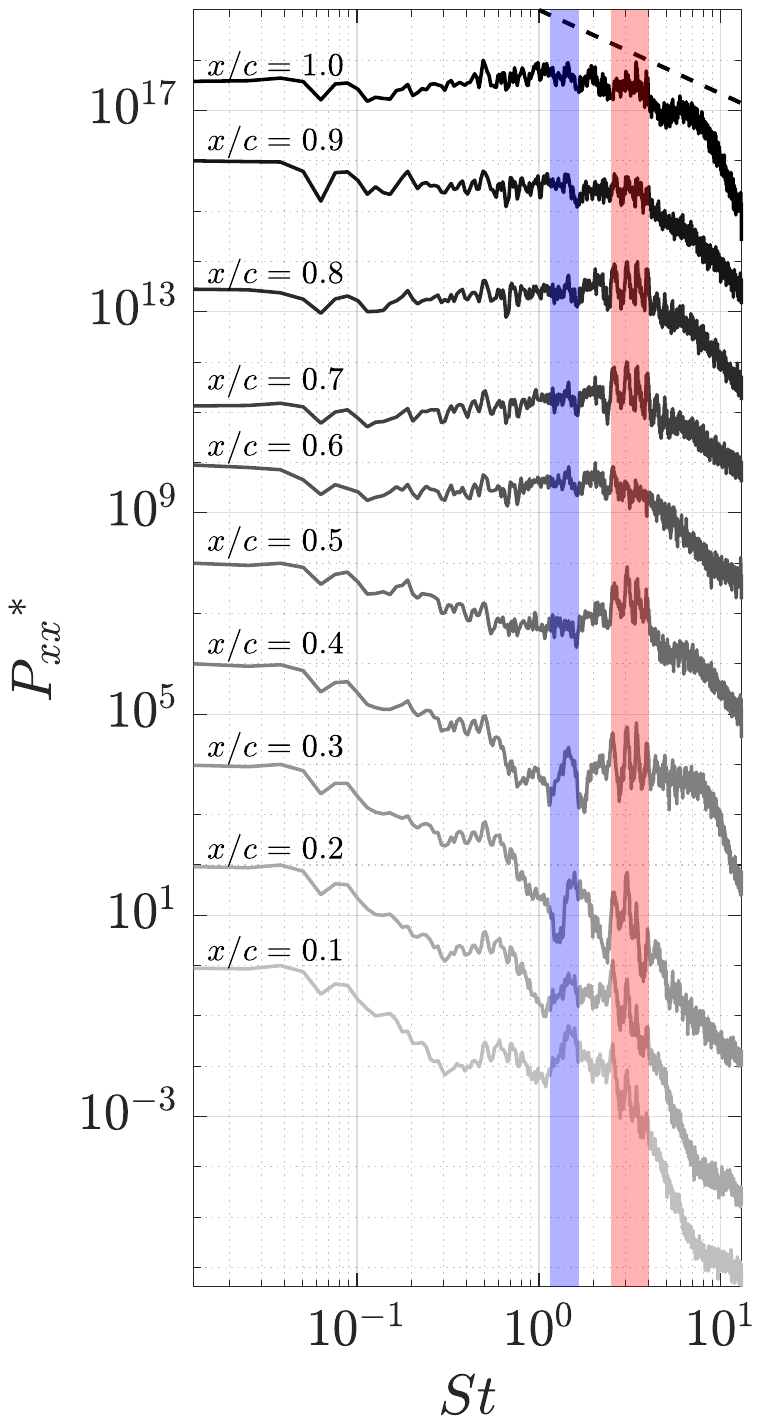}         
         \subcaption[]{${U}'$}
         \label{fig:psd_observables_U} 
    \end{subfigure} 
    \begin{subfigure}[c]{.235\linewidth}
        \centering 
         \includegraphics[trim=4.8cm 1cm 5cm 2.65cm, clip, width =\linewidth]{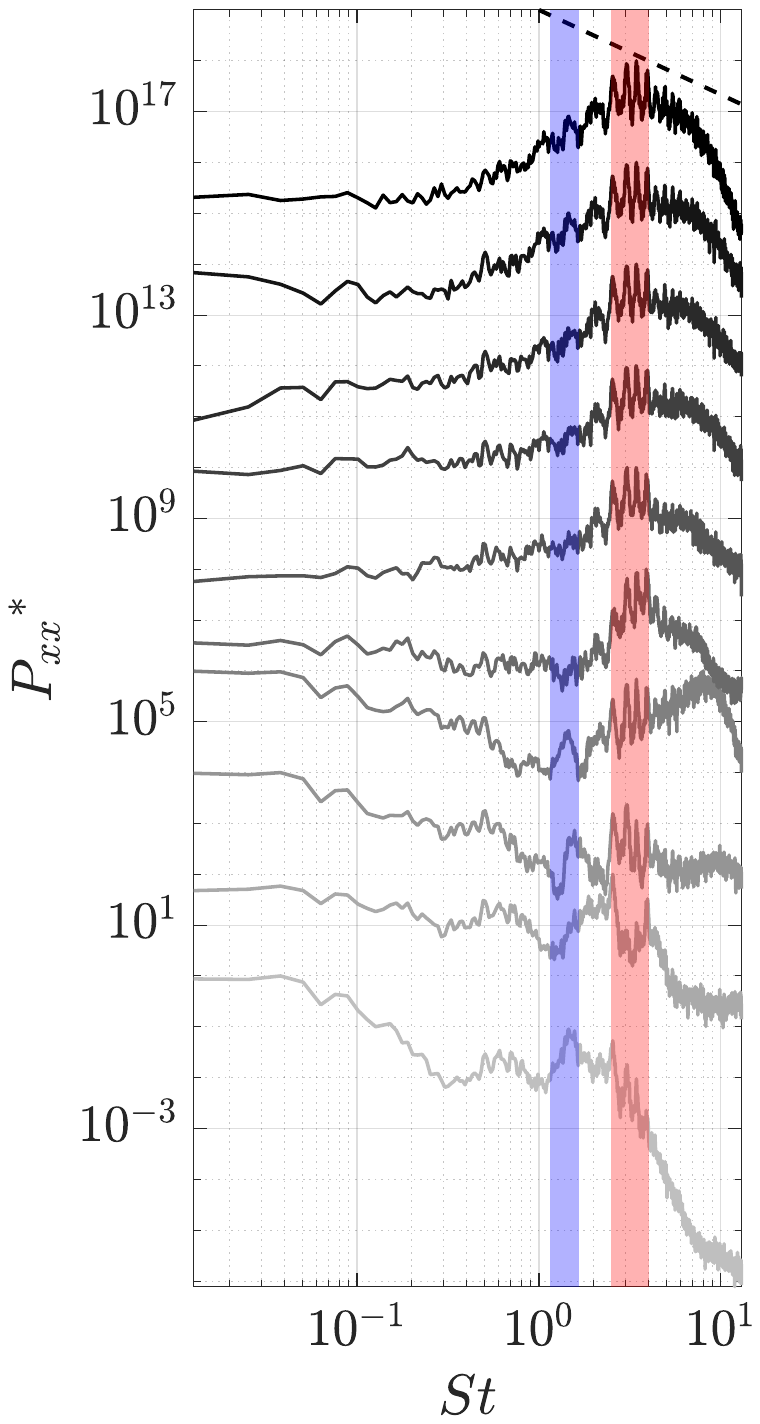}          
         \subcaption[]{${V}'$}
         \label{fig:psd_observables_V}
    \end{subfigure} 
    \begin{subfigure}[c]{.235\linewidth}
        \centering 
         \includegraphics[trim= 4.8cm 1cm 5cm 2.65cm, clip, width =\linewidth]{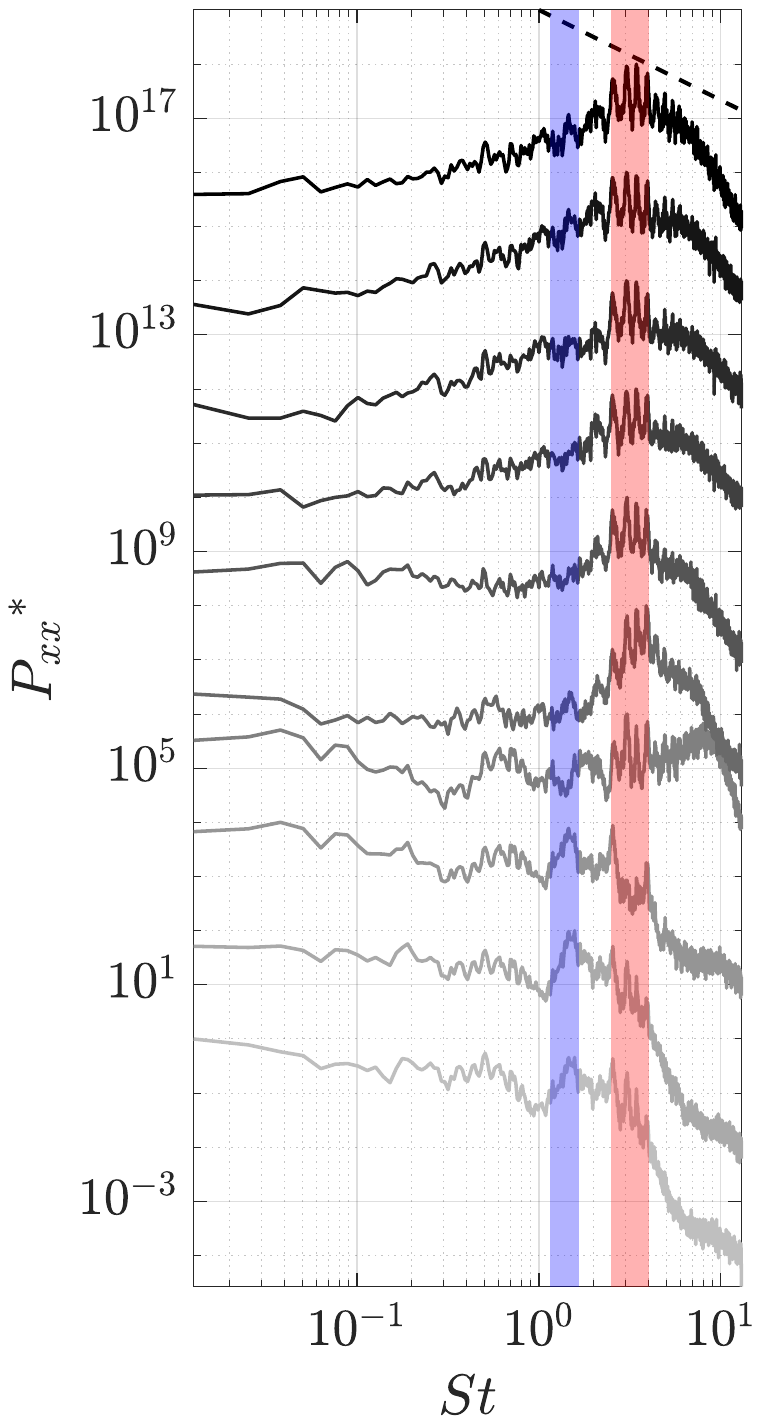}         
         \subcaption[]{$P'$}
          \label{fig:psd_observables_P}
    \end{subfigure} 
    \begin{subfigure}[c]{.235\linewidth}
        \centering 
         \includegraphics[trim= 4.8cm 1cm 5cm 2.65cm, clip, width =\linewidth]{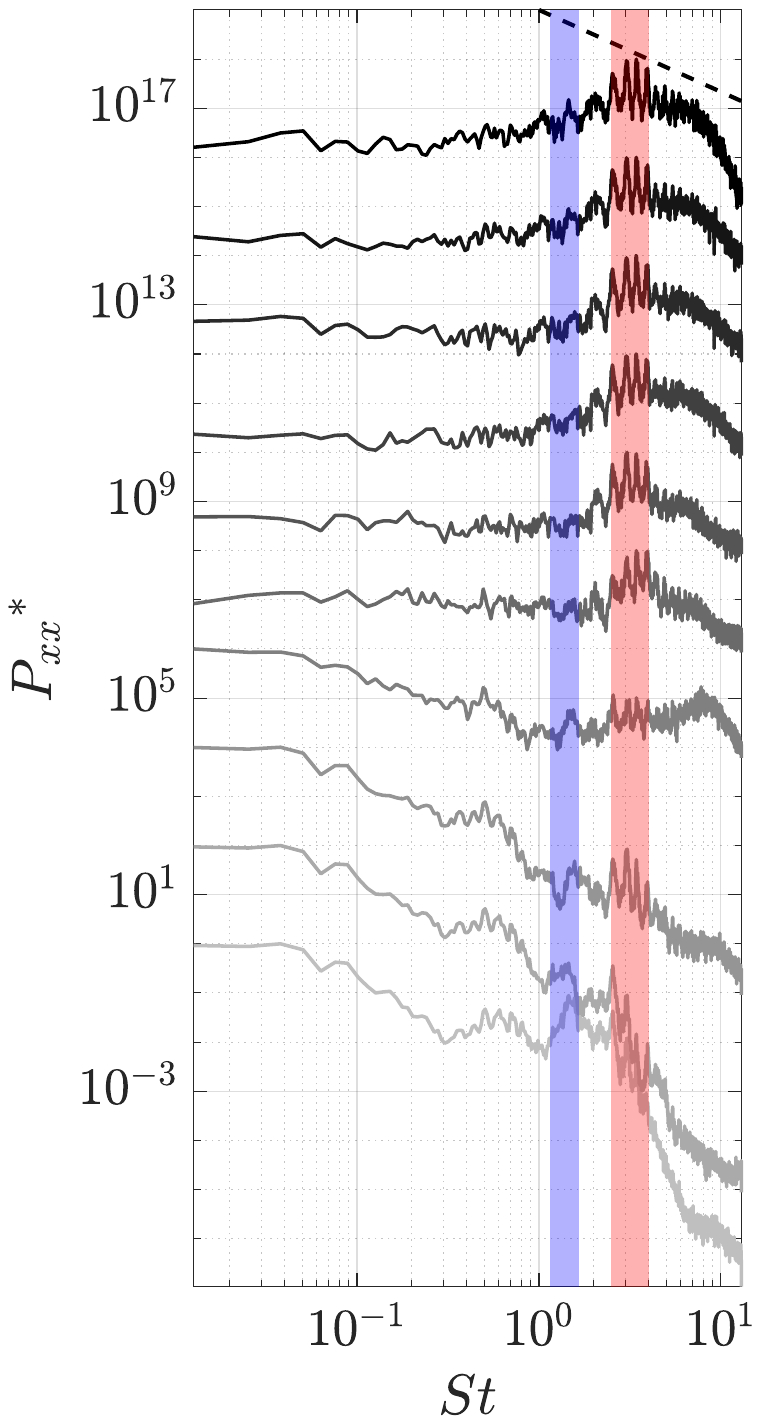}          
         \subcaption[]{${\omega_z}'$}   
         \label{fig:psd_observables_Vort}         
    \end{subfigure} 
\caption{Streamwise development of the unsteady spectral content for
case $C_1$. (a) Locations of probes $P1$--$P10$ from $x/c=0.1$ to
$1.0$. (b--e) Power spectral densities of the fluctuating streamwise
velocity, wall-normal velocity, pressure, and spanwise vorticity,
respectively. Each spectrum is normalized by its maxima and vertically
offset for clarity. The shaded bands identify the upstream
($1.15\lesssim St\lesssim1.65$) and aft-airfoil
($2.5\lesssim St\lesssim4$) frequency ranges discussed in the text.}
    \label{fig:psd_observables}
\end{figure}

To determine whether the spectral content of the aerodynamic loads is associated with spatially coherent flow dynamics, fluctuations in
streamwise velocity, wall-normal velocity, pressure, and spanwise
vorticity are sampled at ten locations extending from the leading-edge
shear layer to the near wake, as shown in
Fig.~\ref{fig:probe_location}. 
The corresponding spectra are shown in Fig.~\ref{fig:psd_observables_U}-\ref{fig:psd_observables_P}. 
Two frequency ranges are highlighted to clarify their streamwise development. 
The blue band, $1.15\lesssim St\lesssim1.65$, identifies spectral content that is most apparent at the upstream probes, whereas the red band, $2.5\lesssim St\lesssim4$, marks the dominant frequency range that develops over the aft portion of the airfoil. 
Each spectrum is normalized by its own maximum and vertically offset for clarity; therefore, the figure is used to compare spectral shapes and dominant frequency ranges rather than absolute fluctuation amplitudes.

At the upstream probes, $P1$--$P3$, the spectra contain broadband fluctuations with noticeable content within the blue band. 
This feature weakens as the separated shear layer evolves downstream and is therefore associated with the early development of shear-layer disturbances. 
Near $P4$, additional higher-frequency content appears, particularly in the wall-normal velocity spectrum near $St\approx8$, coincident with the region of pronounced shear-layer deformation.

Farther downstream, the spectra at $P5$--$P10$ develop a prominent broad band over $2.5\lesssim St\lesssim4$, highlighted in red in
Fig.~\ref{fig:psd_observables_U}-\ref{fig:psd_observables_P}. 
This range appears in the velocity, pressure, and spanwise-vorticity spectra and coincides with the dominant
band observed in the aerodynamic-load spectra. 
Its downstream development therefore associates the red band with vortex formation, interaction, and shedding over the aft portion of the airfoil.

The spectra therefore establish the resolved frequency range of the unsteady flow and connect the broadband aerodynamic loading to the
spatial development of the separated shear layer. They do not, however, identify the transient vortex sequence responsible for an
individual extreme excursion. 
Such a sequence is examined using the event-resolved flow fields in the following sections.

\section{Mechanism of Extreme Drag Events} \label{sec:Mechanism_of_Extreme_Drag_Events}

\subsection{Identification of Extreme Drag Events} \label{sec:EENomenclasture}

\begin{figure}
    \centering    
    \includegraphics[trim= 1.5cm 0cm 3cm 1.2cm, clip, width =0.8\textwidth]{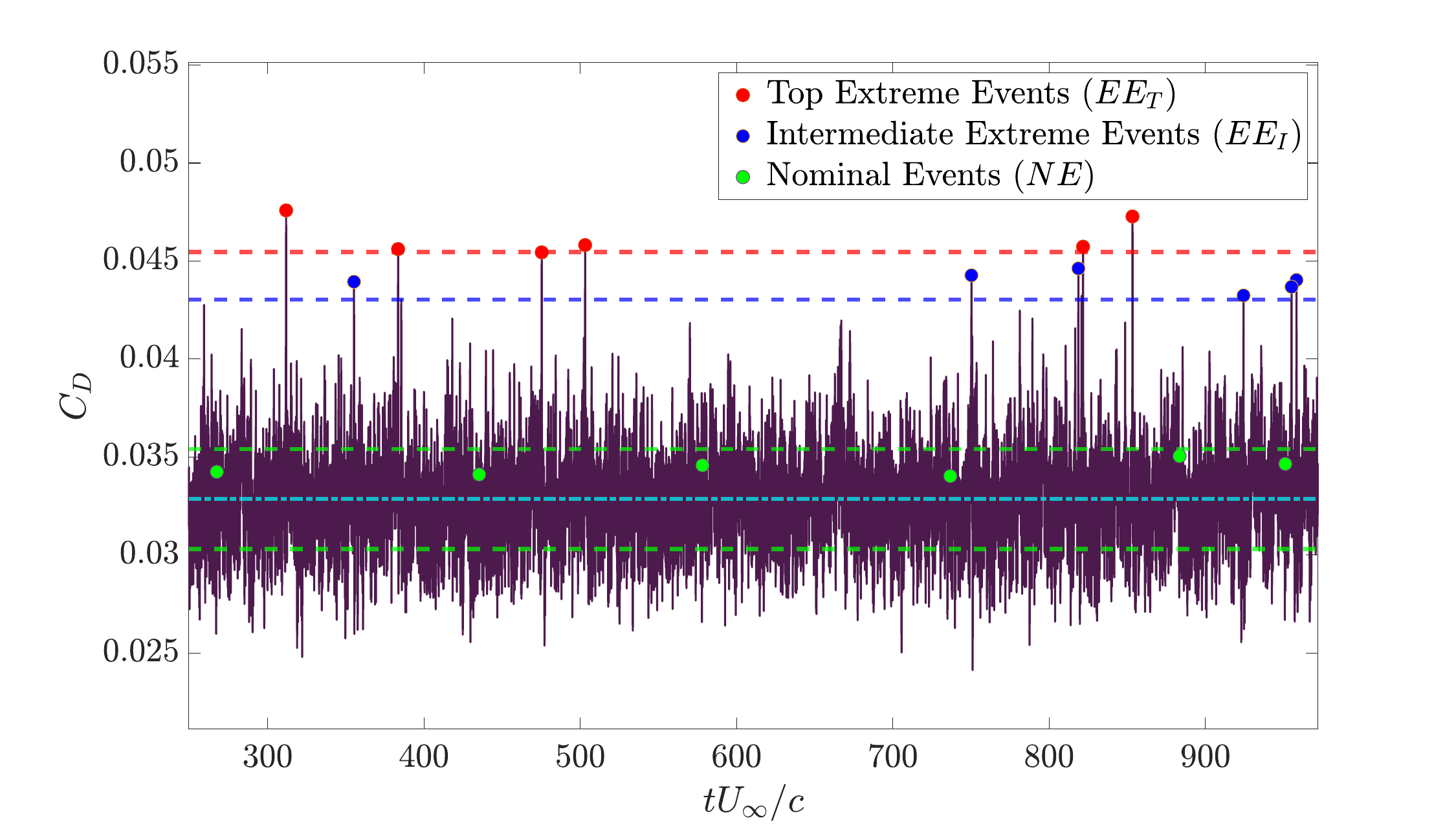}
    \caption{Drag-coefficient history showing the selected top extreme events ($EE_T$), intermediate extreme events ($EE_I$), and nominal events ($NE$). Cyan-colored center line indicates the mean of $C_D$ and other dashed horizontal lines indicate the event-selection thresholds.}
    \label{fig: ee_nee}
\end{figure}

Extreme drag events are identified from the local maxima of the
drag-coefficient history relative to its mean, $\overline{C_D}$, and
standard deviation, $\sigma_{C_D}$. 
Successive threshold crossings belonging to the same drag excursion are treated as a single event and represented by the corresponding local maximum. 
Over the sampled interval of $722c/U_\infty$, the six largest drag excursions exceed $\overline{C_D}+5\sigma_{C_D}$ and are denoted as the top extreme events, $EE_T$. 
The next six largest excursions lie between $\overline{C_D}+4\sigma_{C_D}$ and $\overline{C_D}+5\sigma_{C_D}$ and are denoted as the intermediate extreme events, $EE_I$. 
These 12 events are shown in Fig.~\ref{fig: ee_nee}. Their sparse occurrence within the sampled record highlights the intermittent nature of the largest drag excursions.

For comparison, six representative nominal events, denoted by $NE$, are selected from intervals for which $|C_D-\overline{C_D}|<\sigma_{C_D}$. 
These selection criteria provide consistent extreme and nominal event sets for the event-resolved analysis that follows.

\subsection{Vortex Evolution Leading to Extreme Drag}
\label{sec:Cause and Effect}

\begin{figure}
    \centering
        \begin{subfigure}[c]{\linewidth}
            \centering 
            \includegraphics[trim=0.25cm 0.8cm 4cm 0.2cm, angle =90, clip, width = 1\linewidth]{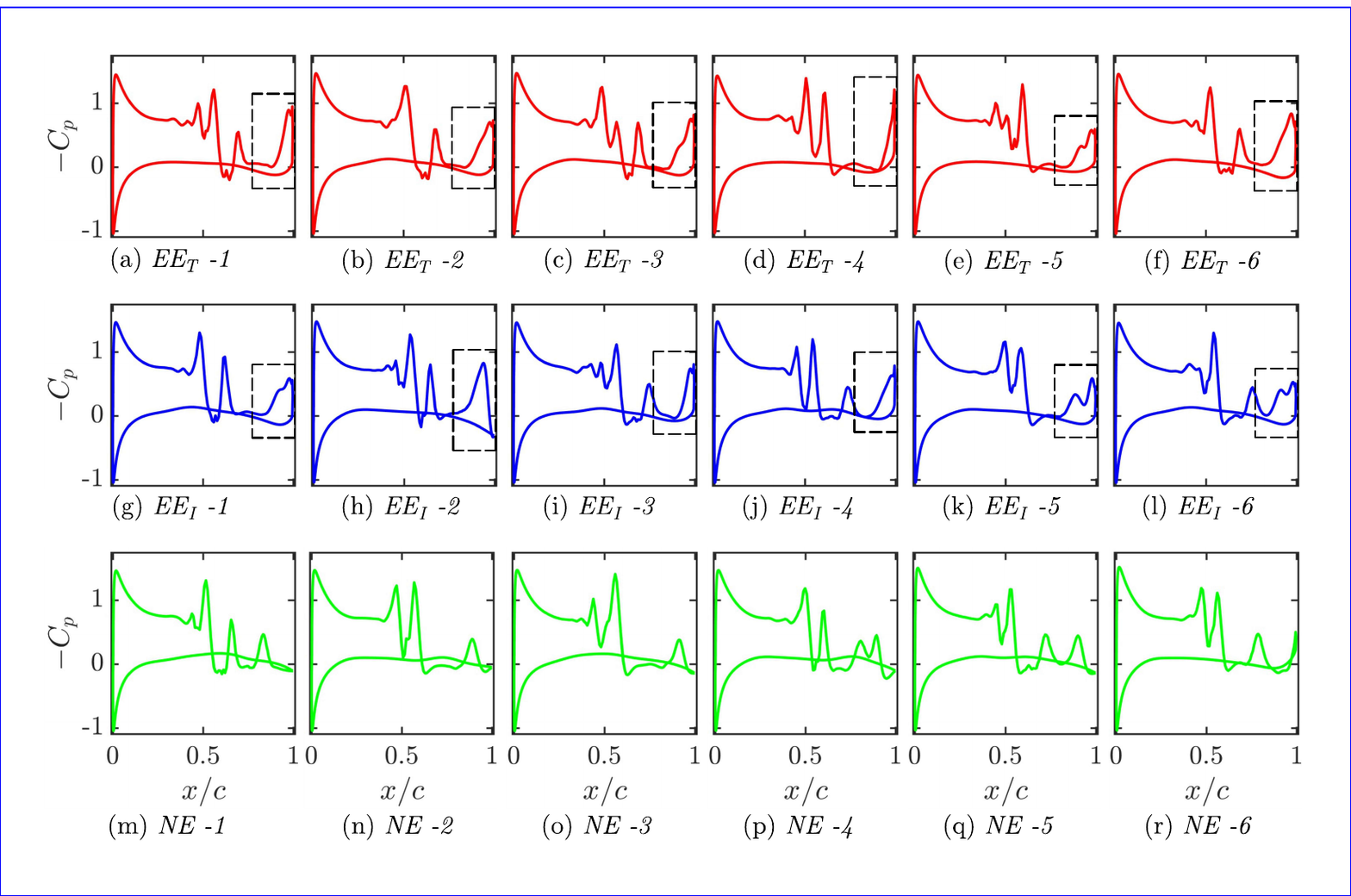}
        \end{subfigure} 
\caption{Instantaneous $C_p$ at the identified $EE_T$ (top row), $EE_I$ (middle row), and
$NE$ (bottom row) cases. The boxes identify the localized trailing-edge suction peaks observed during the extreme events.}
   \label{fig: Inst_Cp_comparison}
\end{figure}

The instantaneous $C_p$ distributions at the selected event
times are compared in Fig.~\ref{fig: Inst_Cp_comparison}. 
All $EEs$ exhibit a pronounced suction peak near the trailing edge on the suction side of the airfoil. 
In contrast, no consistent trailing-edge suction peak is observed among the $NE$. 
The recurrence of this localized pressure signature across all 12 $EEs$ indicates that the extreme drag excursions are accompanied by a distinct change in the unsteady loading near the trailing edge.

The trailing-edge suction peak also  suggests the presence of a coherent
vortical structure passing close to the airfoil surface. 
Examination of the corresponding instantaneous vorticity fields confirms that the pressure minimum coincides with the interaction of vortices near the trailing edge. 
The temporal evolution preceding each event is therefore examined to determine how these structures form and interact.


The instantaneous vorticity field is examined over the 40 stored snapshots preceding each selected event.
An event occurring at $t=t_e$ is represented by 40 preceding flow fields and the event snapshot itself. 
The time corresponding to snapshot
$m$ is
\begin{equation} \label{eq:event}
    t_m=t_e+(m-41)\Delta t,
    \qquad m=1,\ldots,41,
\end{equation}
where $\Delta t$ is the interval between successive stored flow fields and $m=41$ denotes the event time. 

Sequences of vorticity field for each case of $EE_T$, $EE_I$, and $NE$ events are shown in
Fig.~\ref{fig:Low_Re_vort_snap}. 
Four representative times are shown for each sequence, with the final column corresponding to the selected local maximum in $C_D$.
During the $NE$, vortices are shed sequentially from the separated shear layer with sufficient streamwise spacing to convect past the trailing edge without a strong mutual interaction. 
The corresponding $C_D$ signal remains approximately periodic, and no pronounced trailing-edge suction peak develops.

The $EEs$, on the other hand, follow a distinctly different evolution. 
An intermittent bifurcation from the regular shear-layer roll-up produces two or more closely spaced, co-rotating vortices that generally differ in strength.
As these vortices approach the trailing edge, their reduced spacing promotes a strong interaction, consistent with the established
dynamics of co-rotating vortices~\citep{Dritschel1992}. 
At the event time, the vortices often merge over the trailing-edge region on the suction
side of the airfoil. 
The resulting coherent structure produces the
localized pressure minimum identified in
Fig.~\ref{fig: Inst_Cp_comparison} and a simultaneous spike in $C_D$.
The immediate precursor to the extreme event is
therefore not vortex shedding alone, but the interaction of sufficiently strong and closely spaced vortices near the trailing edge.
Similar pairing and merging of suction-side vortices were observed by \citet{ricciardi2022transition}, where the resulting coherent structures retained substantial spanwise coherence as they approached the trailing edge and produced stronger trailing-edge pressure fluctuations and acoustic radiation. 


\begin{figure}
\centering
\begin{minipage}{\linewidth}
    \begin{minipage}[c]{\linewidth}
        \centering
        \includegraphics[trim=0.48cm 0.82cm 16cm 0.93cm, angle=90, clip, width=\linewidth]        {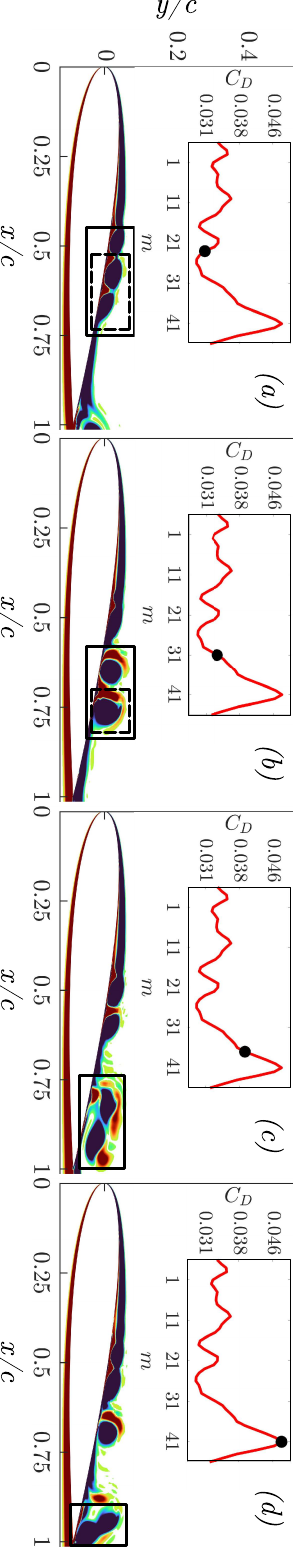}        
    \end{minipage}
\end{minipage}
\vspace{1em}

\begin{minipage}{\linewidth}
    \begin{minipage}[c]{\linewidth}
        \centering
        \includegraphics[trim=0.41cm 0.82cm 16cm 0.93cm, angle=90, clip, width=\linewidth]        {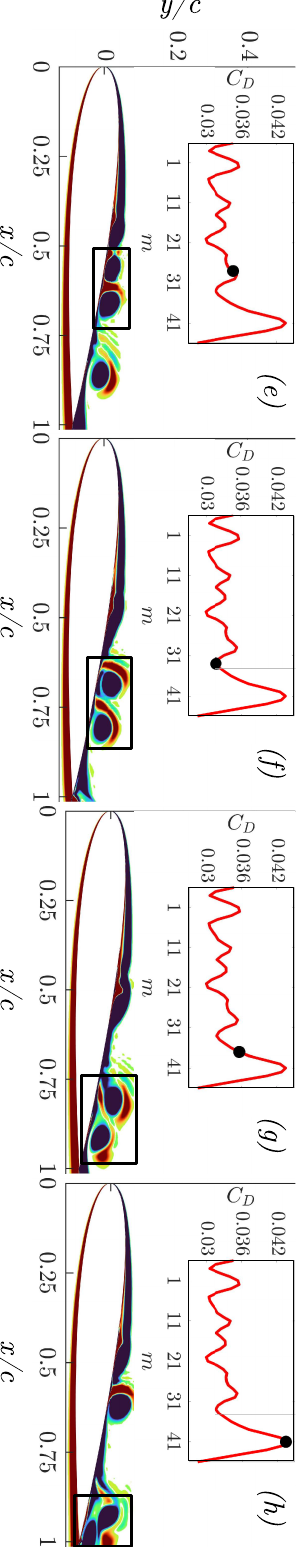}        
    \end{minipage}
\end{minipage}
\vspace{1em}

\begin{minipage}{\linewidth}
    \begin{minipage}[c]{\linewidth}
        \centering
        \includegraphics[trim=0.41cm 0.82cm 16cm 0.93cm, angle=90, clip, width=\linewidth]        {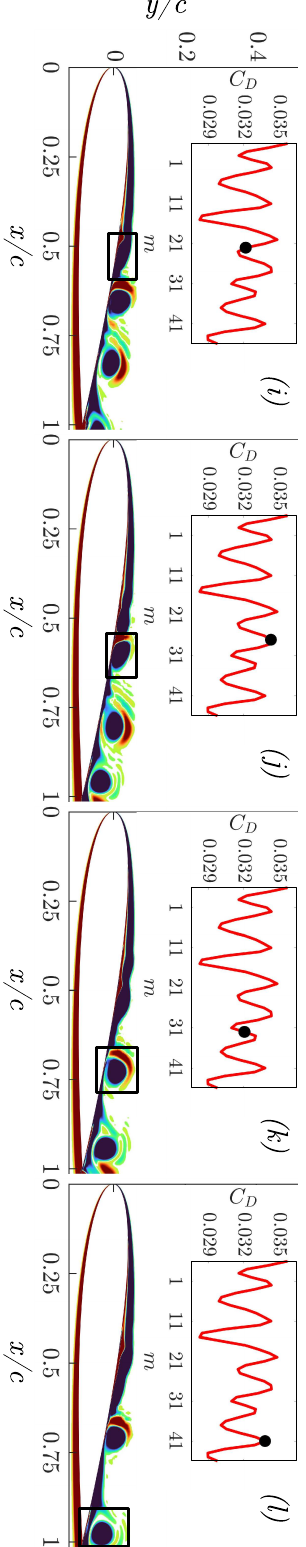}        
    \end{minipage}
\end{minipage}
\vspace{0.5em}

\begin{minipage}{\linewidth}
    \begin{minipage}[c]{\linewidth}
        \centering 
        \includegraphics[trim=22cm 0.1cm 19cm 57cm, clip, width=\linewidth]        {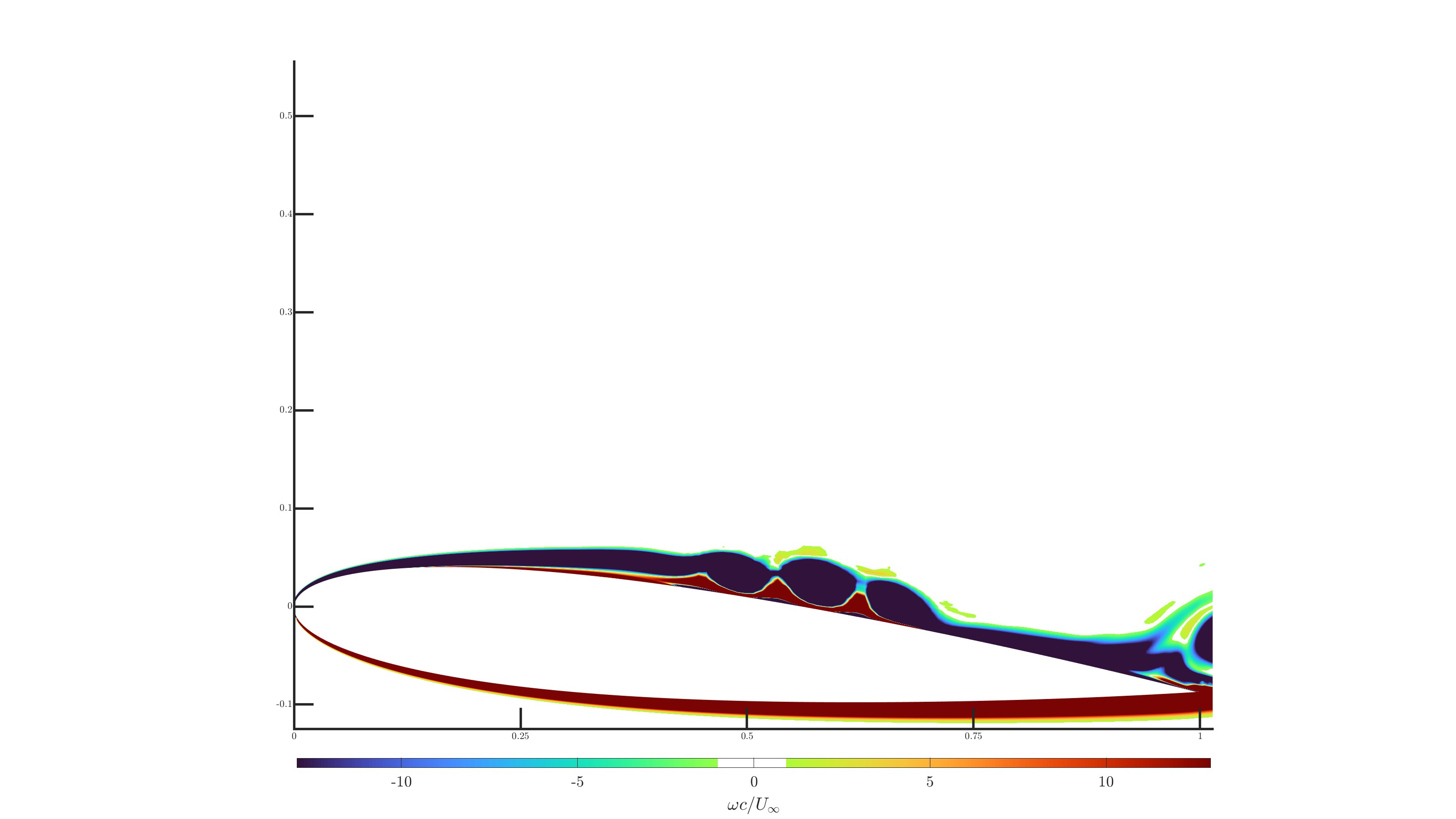}  
    \end{minipage}
\end{minipage}
\label{fig:Low_Re_vort_snap}
\caption{Representative evolution of the instantaneous spanwise
vorticity preceding (a--d) $EE_T$, (e--h) $EE_I$, and (i--l) $NE$
events. Four snapshots are shown for each sequence, with the final
column corresponding to the selected drag maximum at $m=41$. Insets
show the corresponding $C_D$ history and snapshot location within the
pre-event sequence.}
\label{fig:Low_Re_vort_snap}
\end{figure}

It must be noted that vortex merging is not unique to the $EEs$. 
Small-scale mergers occur frequently within the flapping shear-layer region as part of the
nonlinear growth of KH vortices
\citep{Winant_Browand_1974}, but these produce only modest variations in $C_D$. 
The $EEs$ are distinguished by the merger of larger, discrete vortices near the trailing edge, where the resulting pressure signature produces a pronounced spike in $C_D$. 
Complete pre-event sequences for representative $EE_T$, $EE_I$, and $NE$ cases are provided in
supplementary movie~1.

While these observations establish the event-level sequence, they do not explain why the separated shear layer intermittently produces closely
spaced vortices or what controls their subsequent merger. 
The following section therefore quantifies the evolution of the vortices preceding extreme and nominal events.

\subsection{Formation and Release of Closely Spaced Vortices}
\label{sec:Vortex_Characterization}

The preceding section showed that extreme drag events occur when the separated shear layer releases several closely spaced vortices that subsequently interact and merge near the trailing edge. 
To understand how these vortices are produced, we now shift our investigation towards the upstream region, and examine their formation and release within the separated shear layer.




The separated shear layer rolls up into discrete vortices that are eventually shed and propagate towards the trailing edge. 
To distinguish these structures from the opposite-signed vorticity generated near the airfoil surface, they are hereafter referred to as primary vortices (PVs). 
A developing PV remains connected to the upstream portion of the
separated shear layer, hereafter termed the \emph{feeding shear layer},
through which it continues to accumulate vorticity as it grows. The
motion induced by the PV along the airfoil surface, together with the
no-slip condition, generates a thin layer of opposite-signed vorticity
at the wall.

\begin{figure}
    \centering
        \begin{subfigure}[c]{\linewidth}
            \centering
            \includegraphics[trim=0.4cm 0.25cm 6.5cm 0.2cm, clip, angle = 90, width = \linewidth]{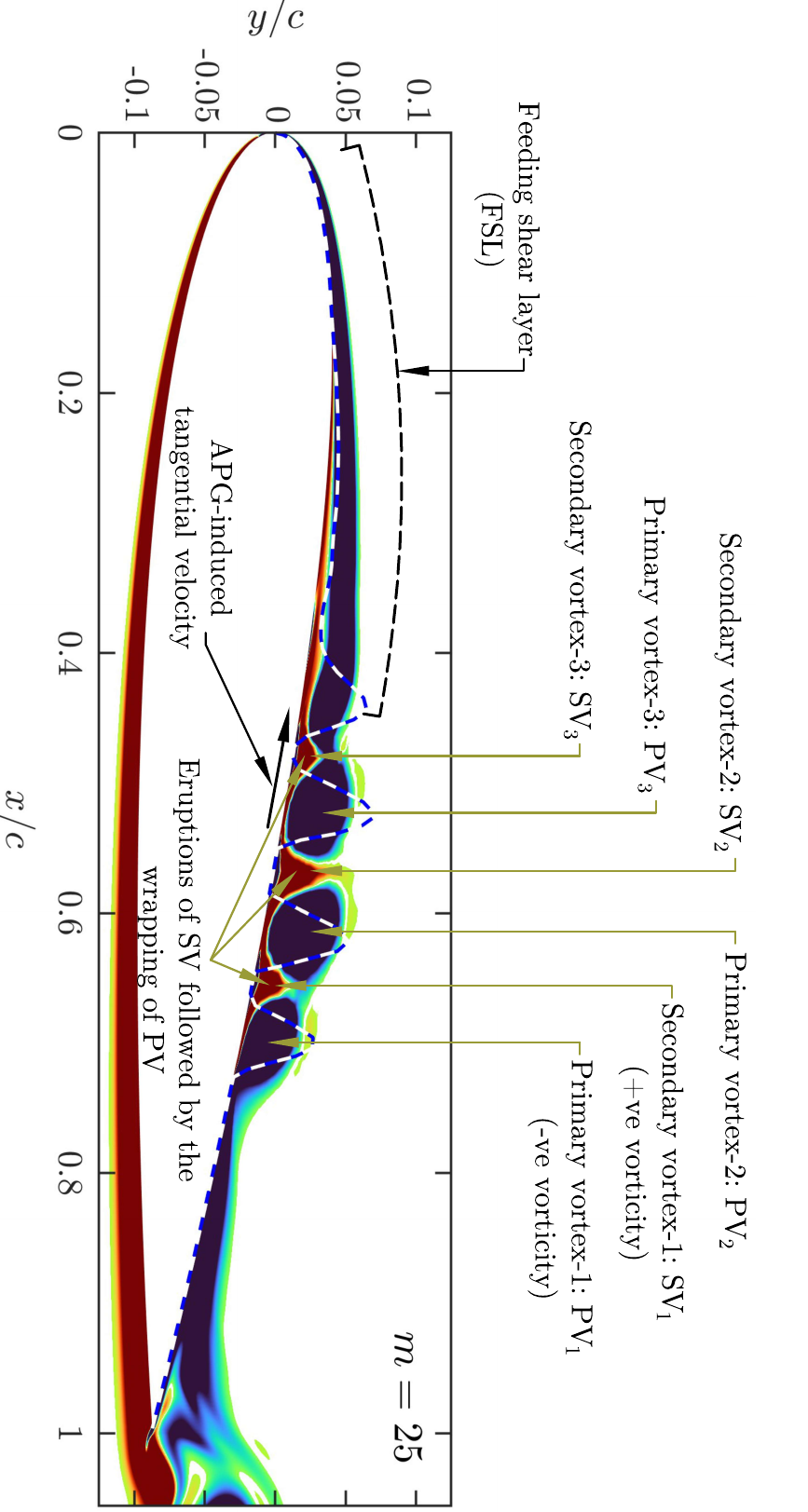} 
        \end{subfigure}
\caption{Instantaneous organization of the vortex-formation region
during $EE_T$--1 at $m=25$. The feeding shear layer, three developing
primary vortices (PVs), and the opposite-signed secondary-vorticity
(SV) layers generated beneath them are identified. The black arrow
indicates the upstream-directed near-wall flow, the dashed line marks
the extent of the reversed-flow region, and the surface-pressure
footprints of the PVs are indicated.}
   \label{fig: Shear_Layer_SV_PV}
\end{figure}

The instantaneous organization of this region is shown in
Fig.~\ref{fig: Shear_Layer_SV_PV} for $EE_T$--1. 
Three developing PVs
are identified together with the feeding shear layer and the
opposite-signed secondary-vorticity (SV) layers beneath them. The
corresponding surface-pressure footprints of the PVs are also
highlighted. As a PV grows, its localized pressure footprint strengthens, producing an increasingly strong APG along the
surface. 
Under the action of this vortex-induced pressure gradient, the SV layer erupts away from the wall. 
Such eruption of wall-generated, opposite-signed vorticity under the
pressure-gradient field imposed by a primary vortex is a fundamental
feature of vortex--wall interactions
\citep{walker1987impact,doligalski1994vortex,
verzicco_orlandi_1996wall,naguib2004wall,prasad2025wavy},
and closely related dynamics have been observed during the evolution and
detachment of unsteady leading-edge vortices
\citep{widmann2015parameters,Vorticity_transport_EoF,
Akkala_Buchholz_2017_Vort_Transport_JFM}. In the present separated shear
layer, the erupting SV penetrates the connection between the PV and its
feeding shear layer, interrupting the continued supply of
negative vorticity and releasing the PV downstream.
Because secondary-vorticity eruption is a viscous response to the
localized pressure field imposed by the PV, the evolution of the
local surface pressure distribution and the streamwise surface-pressure gradient is now examined during the release
process. 

\begin{figure}
\centering 
    \begin{subfigure}[c]{\linewidth}
        \includegraphics[angle=90, trim=0.6cm 0.75cm 2.9cm 1cm,  clip, width=\linewidth]        {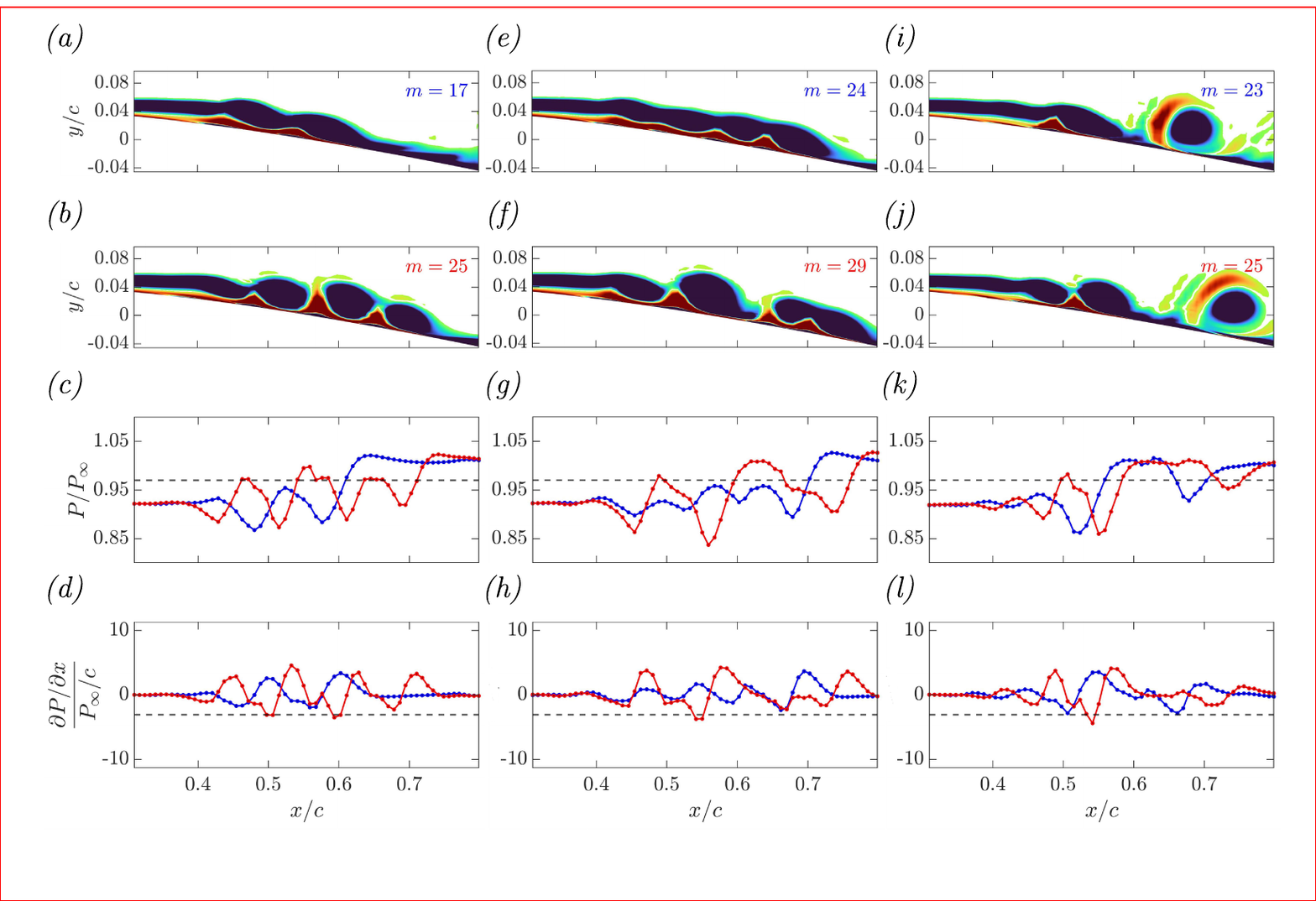}       
    \end{subfigure}
\caption{Surface-pressure evolution associated with secondary-vorticity
eruption and PV release for (a--d) $EE_T$--1, (e--h) $EE_I$--2, and
(i--l) $NE$--2. The first two rows show instantaneous vorticity
immediately before and after SV eruption, respectively. The lower rows
show the corresponding surface pressure, $P/P_\infty$, and normalized
streamwise pressure gradient,
$(\partial P/\partial x)/(P_\infty/c)$. Blue and red curves correspond
to the pre- and post-eruption states.}
    \label{fig:Pressure_SV}
\end{figure}

Figure~\ref{fig:Pressure_SV} examines this process for representative
$EE_T$, $EE_I$, and $NE$ shedding cycles. 
For each case, instantaneous vorticity fields immediately before and after SV eruption are shown together with the corresponding surface-pressure distribution and streamwise pressure gradient. 
Before eruption, a developing PV remains connected to the feeding shear layer, with the wall-generated SV confined beneath it. Following eruption, the SV lifts into the separated flow and interrupts this connection, leaving the PV as a discrete structure that convects downstream.

The accompanying pressure distributions reveal the local forcing
associated with this transition. 
Each PV produces a distinct low-pressure footprint on the surface.
Moving away from the minimum pressure, the surface pressure recovers
toward the freestream value, producing strong streamwise pressure
gradients along the wall.
Because the near-wall flow beneath the PVs is directed upstream
(Fig.~\ref{fig: Shear_Layer_SV_PV}), the negative
$\partial P/\partial x$ branch constitutes an APG relative to the local flow direction. 
As this APG strengthens, the wall-generated SV separates from the surface and erupts into the shear layer.

Although the local release mechanism is common to all three cases, the organization of the surrounding vortices at the time of eruption is markedly different. 
In the $EE_T$ sequence, several PVs occupy the
formation region simultaneously, and eruption produces multiple, same-signed vortices in close succession. 
The $EE_I$ sequence exhibits a similar organization, although with fewer participating structures.
In the $NE$ sequence, by contrast, the released vortex is comparatively isolated from neighboring PVs. 
The pressure distributions reflect the same distinction: the extreme-event sequences contain several closely spaced pressure minima associated with neighboring vortices, whereas the nominal sequence exhibits a more isolated pressure footprint.

Despite these differences in vortex organization, the local pressure
state accompanying SV eruption is remarkably similar. 
Across the shedding cycles examined, eruption occurs as the pressure footprint outside the PV core recovers to approximately $P/P_\infty \simeq 0.97$. A corresponding value of the normalized streamwise pressure gradient, $(\partial P/\partial x)/(P_\infty/c)\simeq -3$, is repeatedly observed within the adverse-gradient region preceding eruption. 
These values are not interpreted as universal critical conditions, but provide a consistent signature of PV release in the present flow.

\begin{figure}
    \centering 
    \begin{subfigure}[c]{\linewidth}  
        \centering 
        \includegraphics[trim=1.2cm 0.8cm 7.5cm 0.54cm, angle = 90, clip, width=\linewidth]{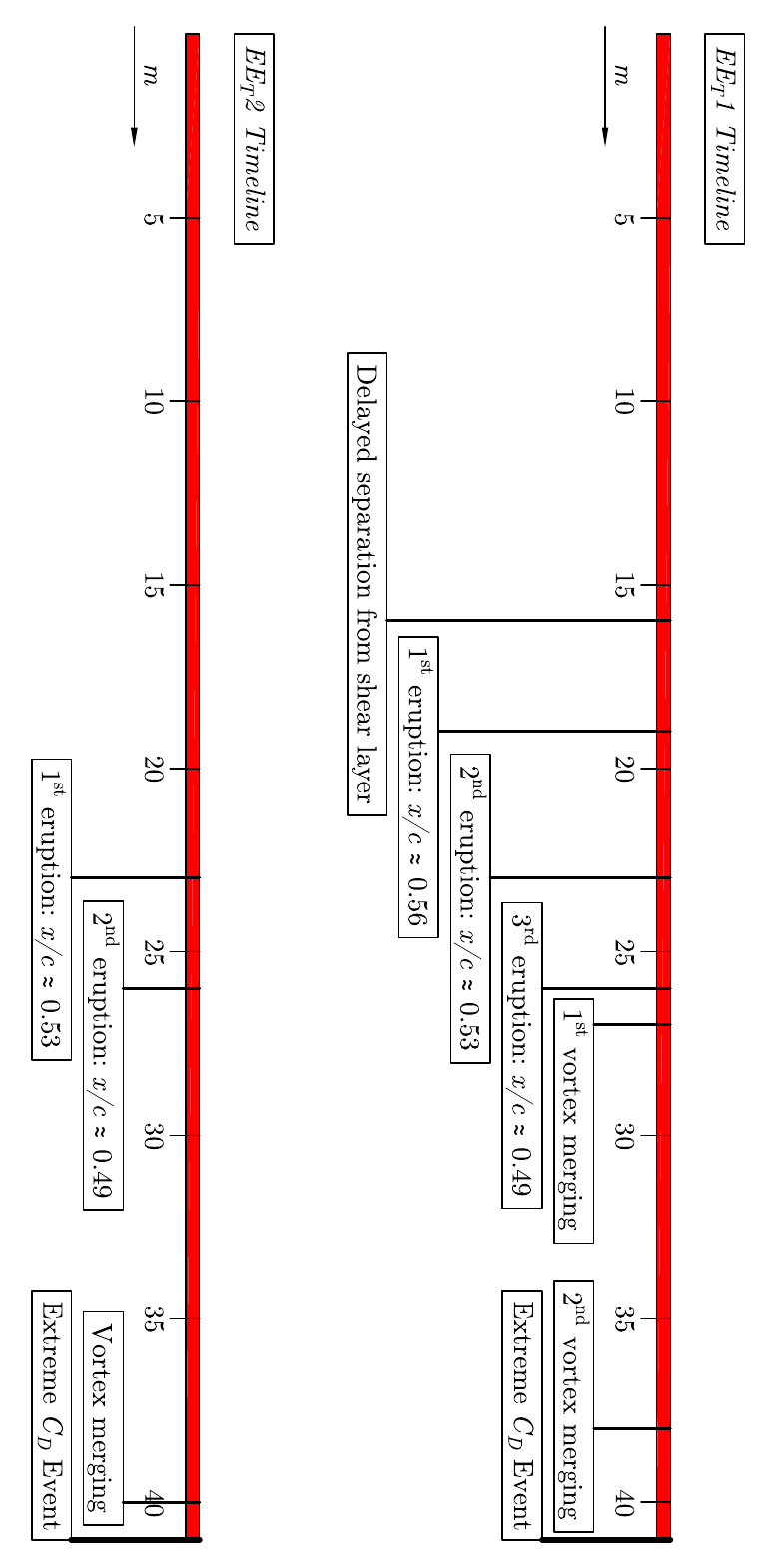}
        \subcaption{}
        \label{fig: EE_Timeline} 
    \end{subfigure}
    \begin{subfigure}[c]{\linewidth}  
        \centering %
        \includegraphics[trim=1.2cm 0.8cm 11.7cm 1cm, angle = 90, clip, width=\linewidth]{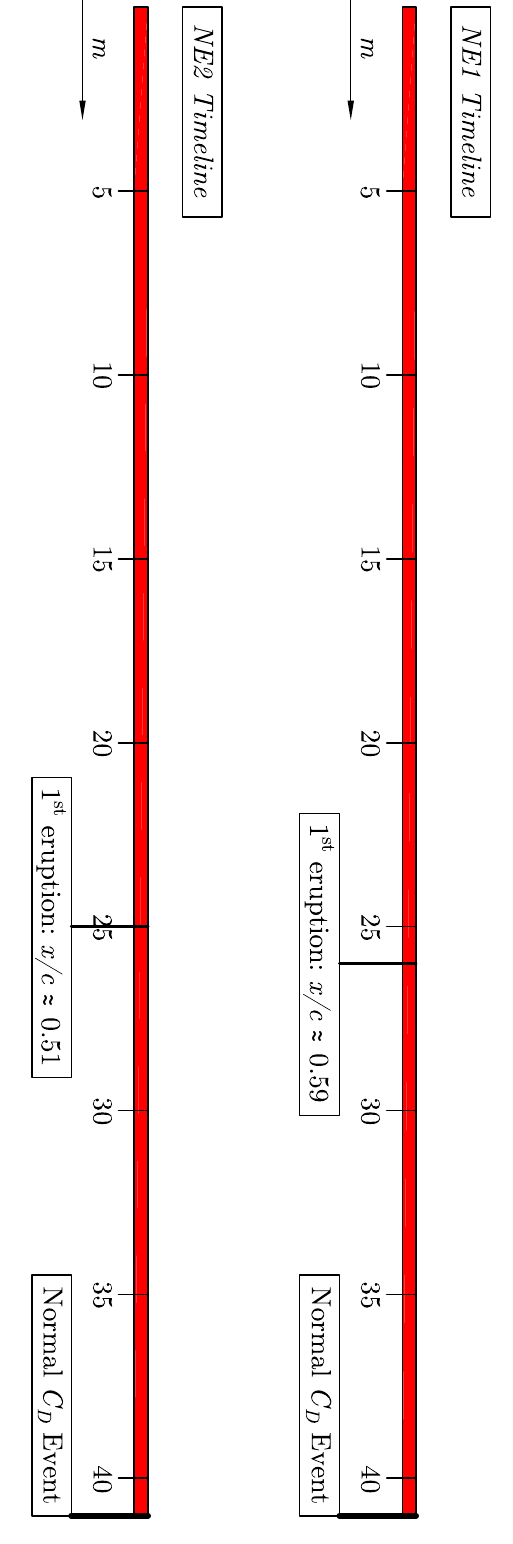}
        \subcaption{}
        \label{fig: NE_Timeline} 
    \end{subfigure}            
\caption{Timelines of the secondary-vorticity eruptions associated with the vortex or vortex system reaching the trailing-edge region at
$m=41$: (a) $EE_T$--1 and $EE_T$--2 and (b) $NE$--1 and
$NE$--2.}
    \label{fig:EE_NE_Timeline}    
\end{figure}

The instantaneous fields therefore suggest that the distinction between extreme and nominal shedding lies not in how an individual PV is released, but in how several such releases are organized. To examine this directly, the vortex or vortex system occupying the trailing-edge region at the event time is traced backward through the pre-event sequence to its release from the feeding shear layer. The resulting release histories are summarized in Fig.~\ref{fig:EE_NE_Timeline}.

For the $EE_T$ cases, the trailing-edge vortex system originates from
several distinct PVs released by SV eruptions occurring within a
relatively short interval. 
In $EE_T$--1, three eruptions release the
vortices that subsequently undergo two successive interactions, whereas $EE_T$--2 contains two closely spaced eruptions preceding the final merger. 
Successive PVs are therefore released before the preceding structures have convected far downstream, establishing the small
streamwise spacing required for their subsequent interaction.

The nominal events exhibit a fundamentally different release history.
For each $NE$ shown in Fig.~\ref{fig:EE_NE_Timeline}, the vortex that
subsequently reaches the trailing-edge region is associated with a
single SV eruption. No additional PV is released sufficiently close in time to accompany it downstream, and the vortex consequently approaches the trailing edge as an isolated structure rather than as part of the compact multi-vortex system observed during the extreme events.

SV eruption therefore provides the local mechanism by which an individual PV is released from the feeding shear layer, while the spatial and temporal organization of successive eruptions determines the spacing of the released vortices. 
The clustered eruption sequence observed before the $EE_T$ states provides a potential
precursor to the later trailing-edge merger and identifies the
vortex-release region as a possible target for modifying the downstream
vortex interactions and aerodynamic loading.

\subsection{Circulation and Convection of the Interacting Vortices} \label{sec:circulation_tracking}

We now quantify the circulation and convection of the interacting PVs to determine whether their merger is associated with a consistent ordering of vortex strength or downstream speed. 
The vortex centers and boundaries are identified using the gradient-free $\Gamma_1$ and
$\Gamma_2$ functions introduced by \citet{Graftieaux_2001}. 
This approach has been widely applied to identify and characterize coherent vortices in experimental and numerical aerodynamic flows~\citep{rod_airfoil_ther, Baik_Bernal_Granlund_Ol_2012,
Pitt_Ford_Babinsky_2013,biesinger2024effect}. 
Local extrema of $\Gamma_1$ locate the vortex centers, while the contour $|\Gamma_2|= 2/\pi$ defines their boundaries. 
The minimum value of $|\Gamma_1|$ and $|\Gamma_2|$ used to retain candidate centers and boundaries, respectively,
varies among implementations~\citep{Baik_Bernal_Granlund_Ol_2012, Pitt_Ford_Babinsky_2013,beverly,simpson2018detecting,carvajal_gamma2, milner2025_magnus,yadala2025effect,mcatee2026_vortex_ring_collisions}.
Here, a permissive threshold of
$|\Gamma_1|\geq0.3$ is used to retain weakly developing and strongly
deformed PVs. 
Candidate vortices are then associated across successive snapshots using continuity of their position and vorticity sign, while severely isolated detections that do not form a continuous trajectory are
discarded.
The reported trajectories and circulation trends are insensitive to
moderate variations in this candidate-detection threshold.

The circulation and streamwise convection speed are evaluated as
\begin{equation}
    \Gamma(m)
    =
    \iint_{\mathcal{A}_v(m)}\omega_z\,dA,
    \qquad
    U_c(m)=\frac{dx_c}{dt},
    \label{eq:vortex_tracking_quantities}
\end{equation}
where $\mathcal{A}_v(m)$ is the region enclosed by the corresponding $|\Gamma_2|=2/\pi$ contour and $x_c$ is the streamwise location of the vortex center. 
Vortex strength and convection speed are reported as $\Gamma^*=|\Gamma|/(cU_\infty)$ and $U_c^*=U_c/U_\infty$, respectively.

The current implementation of the $\Gamma_1$--$\Gamma_2$ method is
formulated on a uniform Cartesian sampling grid. 
Accordingly, the  velocity field from the body-fitted curvilinear CFD mesh is interpolated onto a Cartesian grid containing $601\times251$ points over the
suction-side region spanning across $0.35\lesssim x/c\leq1$. 
Further refinement of the interpolated grid produces negligible changes in the detected vortex
centers, circulation histories, and convection speeds.

\begin{figure}
    \centering 
    \begin{subfigure}[c]{\linewidth}  
            \centering 
            \includegraphics[trim=1.35cm 4.1cm 1cm 1.35cm, angle = 90, clip, width=\linewidth]{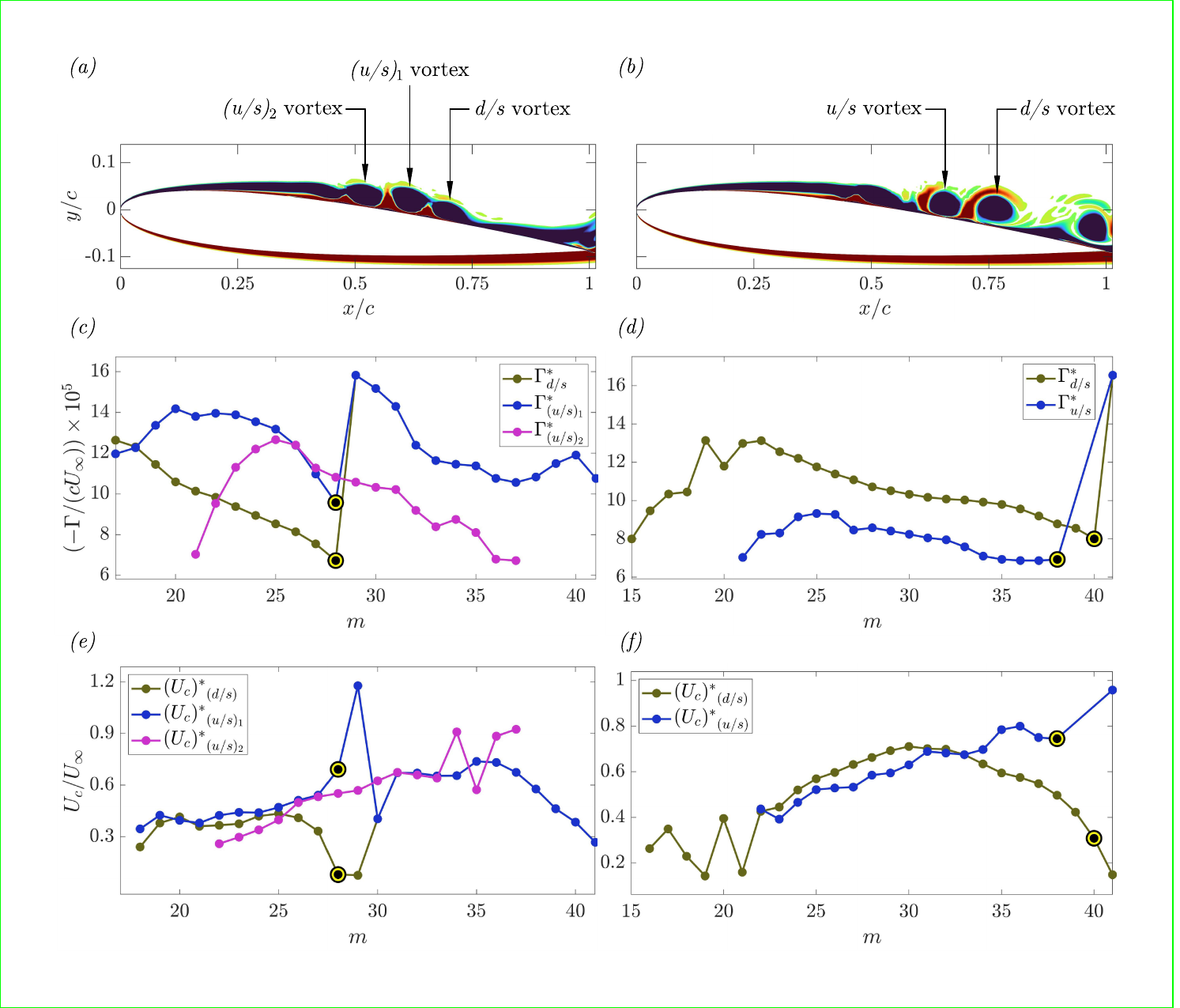}
        \end{subfigure} 
\caption{Evolution of the interacting vortices during
$EE_T$--1 (left) and $EE_I$--1 (right).
(a,b) Instantaneous vorticity fields with the tracked downstream
($d/s$) and upstream ($u/s$) vortices identified.
(c,d) Normalized vortex circulation, $\Gamma^*$.
(e,f) Normalized streamwise convection speed, $U_c^*$.
For $EE_T$--1, $u/s_1$ and $u/s_2$ denote the first and second
upstream vortices participating in successive interactions.}
    \label{fig:Circulation_Tracking_Total}    
\end{figure}

Using this framework, the interacting vortices are tracked through two representative extreme events. 
Figures~\ref{fig:Circulation_Tracking_Total}(a)
and (b) show the instantaneous vorticity fields for $EE_T$--1 and
$EE_I$--1, respectively, together with the nomenclature used to identify
the downstream vortex, $d/s$, and the successive upstream vortices,
$u/s_1$ and $u/s_2$. 
Their circulation histories are shown in
Fig.~\ref{fig:Circulation_Tracking_Total}(c) and (d).
The $EE_T$--1 sequence contains two successive interactions. 
During the first, $u/s_1$ approaches the downstream vortex and the two detected
cores coalesce shortly after $m=28$. 
Immediately before coalescence, the sum of their circulation magnitudes is
$(\Gamma_{d/s}^*+\Gamma_{u/s_1}^*)=16.296 \times 10^{-5}$, whereas the circulation
of the resulting vortex is $\Gamma^*=15.823 \times 10^{-5}$ at $m=29$. 
The difference is approximately $3\,\%$, indicating that most of the
circulation contained within the two detected regions is recovered in
the merged structure.

The resulting downstream vortex subsequently interacts with $u/s_2$ near the trailing edge. 
Over approximately $m=37$--$40$, the circulation associated with $u/s_2$ decreases while that of the downstream structure increases. 
This complementary evolution is consistent with vorticity from the weaker upstream structure being stretched and incorporated into the larger downstream vortex. 
The interaction therefore appears to proceed through a
straining- and filamentation-dominated stage rather than immediate
coalescence of the two cores
\citep{Dritschel1992,Hopfinger_Heijst_1993,
amoretti2001asymmetric,CERRETELLI_WILLIAMSON_2003,
brandt2007physics,leweke2016dynamics_AR}.
The two structures remain distinguishable at $m=41$, but their
interaction continues downstream and ultimately leads to coalescence.

The $EE_I$--1 sequence, on the other hand, contains a single interaction. 
In this case, the two detected cores approach and coalesce into a single vortex near $m=41$, producing the sharp change in the circulation history. 
The pre-merger circulation sum is $(\Gamma_{d/s}^*+\Gamma_{u/s}^*)=14.926 \times 10^{-5}$, compared with
$\Gamma^*=16.547 \times 10^{-5}$ for the final detected structure. 
The approximately $10\,\%$ difference should not be interpreted as a
physical production of circulation because the upstream vortex is not
reliably detected at $m=39$. It is more likely to reflect changes in the
$\Gamma_2$ integration boundary during the strongly deformed
interaction.

The circulation histories also show that merger does not require a
fixed streamwise ordering of vortex strength. 
During the first interaction in $EE_T$--1, the upstream vortex $u/s_1$ is stronger than the downstream vortex. 
However, during the second interaction in $EE_T$--1, $u/s_2$ is weaker than the downstream structure. 
In addition, the upstream vortex is also weaker in $EE_I$--1. 
Thus, a stronger upstream vortex is not a necessary condition for the interaction to progress toward merger.

The convection-speed histories in Figs.~\ref{fig:Circulation_Tracking_Total}(e) and (f) show that the relative motion of the vortices develops as each interaction proceeds.
Before the first merger in $EE_T$--1, $u/s_1$ accelerates while the downstream vortex decelerates sharply, producing a rapidly increasing speed differential as the two cores approach. 
The abrupt change in the speed of the downstream structure immediately after $m=28$ is not interpreted, because the merger replaces two detected centers with the center of a newly formed vortex.

During the subsequent interaction between this merged downstream structure and $u/s_2$, the two vortices initially convect at comparable speeds. 
As the interaction develops, however, $u/s_2$ generally convects faster, while the downstream structure slows toward the end of the sequence. 
Although the histories fluctuate because both vortices are strongly deforming, the resulting speed differential continues to reduce their streamwise separation before their eventual merger downstream of $m=41$.

A similar evolution occurs in $EE_I$--1. 
Both vortices initially accelerate, but the downstream vortex subsequently slows while the upstream vortex maintains or further increases its convection speed.
Thus, the speed differential develops through somewhat different temporal pathways in the three interactions, but in each case the upstream vortex ultimately gains on the downstream structure. 
The release of clustered structures identified in $\S~\ref{sec:Cause and Effect}$ establishes the initially small vortex spacing, while the subsequent differential convection promotes their interaction near the trailing edge.

The interactions examined in Fig.~\ref{fig:Circulation_Tracking_Total} include both rapid coalescence and a more prolonged deformation of the participating
vortices. 
In particular, the second interaction in $EE_T$--1 exhibits a decrease in the circulation of the upstream vortex accompanied by an increase in that of the downstream structure, consistent with stretching and redistribution of vorticity before the vortices eventually coalesce farther downstream.

\begin{figure}
    \centering 
    \begin{subfigure}[t]{\linewidth}  
            \centering 
            \includegraphics[trim=1.45cm 1.6cm 8.5cm 1.75cm, angle = 90, clip, width=\linewidth]{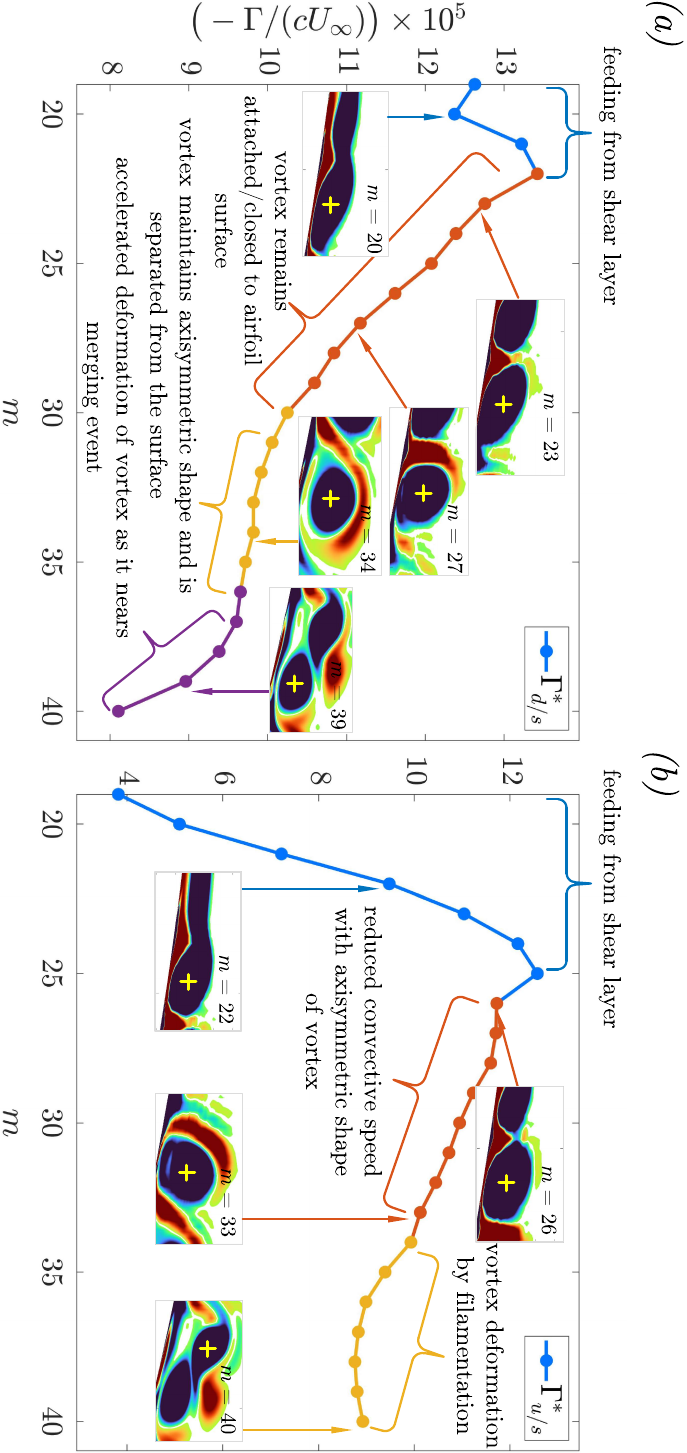}
            \label{fig: Circulation_Trend_Characterization_A} 
    \end{subfigure}   \vspace{-0.4cm}
\caption{Circulation evolution of the interacting vortices during
$EE_I$--6. (a) Downstream and (b) upstream PVs through the
feeding-and-growth, post-release, and interaction-and-deformation
phases. Insets show representative instantaneous vorticity fields;
crosses mark the tracked vortex centers.}
    \label{fig:Circulation_Trend_Characterization} 
\end{figure}

To examine this behavior more clearly, Fig.~\ref{fig:Circulation_Trend_Characterization} considers
$EE_I$--6. 
In this event, the two vortices remain separately identifiable throughout the available pre-event sequence and do not undergo complete coalescence by $m=41$. 
The event therefore isolates the extended interaction and deformation phase without the abrupt changes in vortex identity, associated with complete merger. The circulation histories and corresponding vorticity fields reveal three
successive phases in the evolution of the tracked vortices.

During the first, termed as \textit{feeding-and-growth phase}, each PV
remains connected to the feeding shear layer and its circulation
increases as primary-signed vorticity is incorporated. 
The circulation reaches a maximum near the time at which the connection to the feeding shear layer is interrupted, and the vortex is released downstream.
This is followed by a \textit{post-release phase}. 
Once the vorticity supply is interrupted, the circulation decreases gradually as the vortex convects over the aft portion of the airfoil. 
During this interval, the vortex remains relatively compact and undergoes only modest deformation.

The final \textit{interaction-and-deformation phase} begins as the two
vortices approach one another. 
The structures become increasingly distorted, but remain separately identifiable through $m=41$. 
The circulation of the downstream vortex decreases rapidly, whereas the
upstream vortex exhibits a modest increase near the end of the
sequence. 
Together with the stretching visible in the instantaneous vorticity fields, these complementary trends are consistent with filamentation and redistribution of vorticity during a prolonged interaction. 
This evolution resembles the second interaction in $EE_T$--1, but is more clearly isolated here because complete coalescence does not occur within the investigated interval.


The durations and detailed circulation histories of these phases vary
among events. 
Nevertheless, $EE_I$--6 provides a clear example of the extended-deformation pathway that complements the rapid-coalescence events examined previously: circulation growth while connected to the feeding shear layer, gradual weakening following the release, and prolonged filamentation during vortex interaction. 
Importantly, the drag maximum can occur before the two detected cores have completely coalesced. 
The aerodynamic response is therefore associated with the strong near-trailing-edge interaction and deformation of the vortex system, rather than strictly requiring the completion of the merger at the precise event time.

\section{Effect of Reynolds Number and Three-Dimensional Implications}
\label{sec:High_Re}
The mechanism identified in $\S$~\ref{sec:Mechanism_of_Extreme_Drag_Events}
was established using case $C_1$ at $Re=50{,}000$. We now examine case $C_2$ to determine whether the same event pathway remains active after an order of magnitude increase in $Re$. 
The flow conditions and numerical configuration of $C_2$ were introduced in
$\S~\ref{sec: DNS}$; only its event-resolved dynamics are considered here.

\subsection{Persistence of Vortex Merging}

Although case $C_2$ is two-dimensional, the wall-resolved calculation remains computationally demanding. 
The thinner boundary layer and smaller vortical structures at $Re=500{,}000$ require substantially finer spatial and temporal resolution than in case $C_1$. 
Consequently, the sampled record for $C_2$ is shorter than that available for $C_1$. 
The purpose of this calculation is not to repeat the complete mechanism-resolving analysis performed at $Re=50{,}000$, but to address two fundamental questions: whether pronounced intermittent drag excursions remain persistent at the higher $Re$, and whether the smaller vortices generated under these conditions can still organize into structures that produce an appreciable response in $C_D$.

\begin{figure}
    \centering
        \begin{subfigure}[c]{\linewidth}
            \centering 
            \includegraphics[angle=90,trim=0.78cm 0.75cm 10.8cm 0.67cm, clip, width=\linewidth]{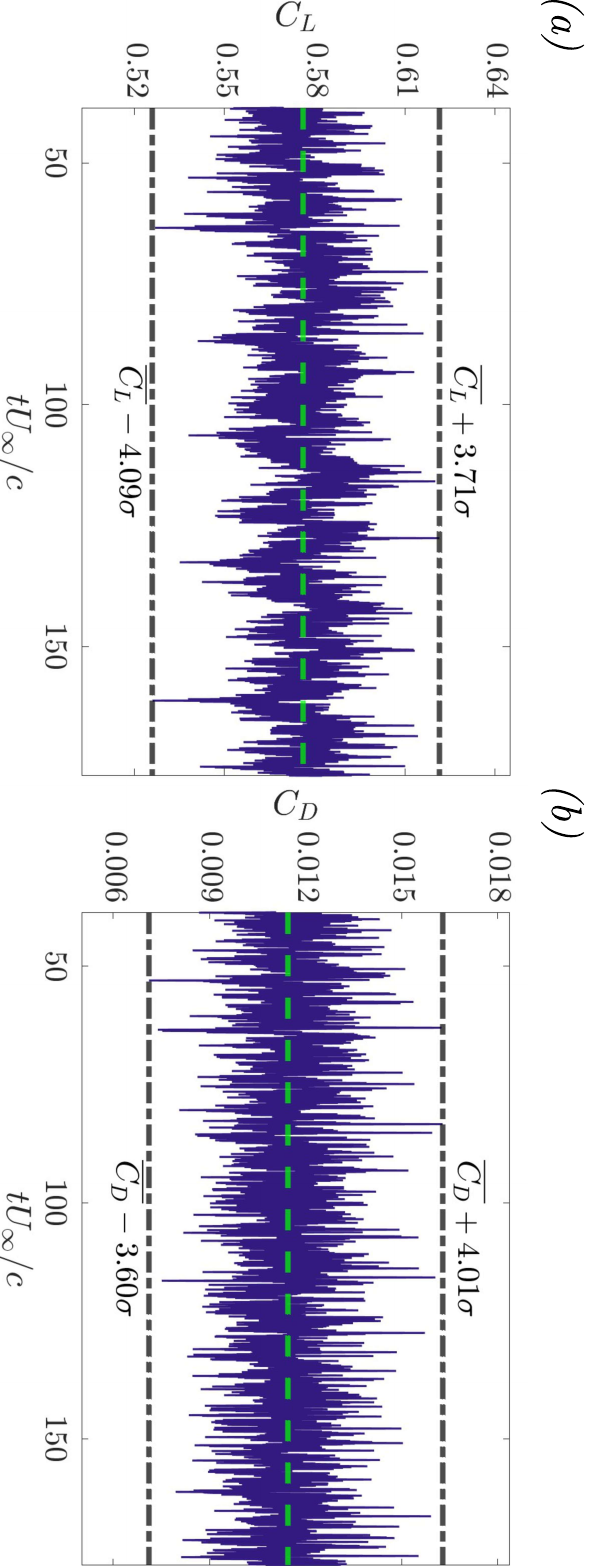}\label{fig: 1}
        \end{subfigure}
 \caption{Time histories of (a) $C_L$ and (b) $C_D$ for the
$Re=500{,}000$ case ($C_2$). Dashed lines denote the mean values and
dash-dotted lines denote the extrema over the sampled interval.}
    \label{fig: high_Re_Time_Signal}    
\end{figure}

The aerodynamic-load histories for $C_2$ are shown in Fig.~\ref{fig: high_Re_Time_Signal}. Both $C_L$ and $C_D$ exhibit irregular fluctuations, and the $C_D$ history contains several isolated positive excursions that rise well above the surrounding variations.
Thus, despite the smaller characteristic scale of the vortical structures at the higher $Re$, pronounced transient drag excursions remain present.

The largest excursion in the sampled $C_2$ record reaches
$\overline{C_D}+4.01\sigma_{C_D}$, compared with
$\overline{C_D}+5.78\sigma_{C_D}$ for the strongest event in the longer $C_1$ record. 
These values are reported only to characterize the available signals and should not be interpreted as evidence that event
severity decreases with increasing $Re$. 
The maximum observed excursion is strongly dependent on the duration of the sampled record, and the shorter $C_2$ record does not provide a statistically converged estimate of either the distribution tail or the event-occurrence rate.

The three largest distinct local maxima in the $C_D$ history are selected for subsequent event-resolved examination.
These cases are used to determine whether the vortex organization accompanying the largest excursions in $C_2$ resembles the extreme-event pathway identified for $C_1$.
Each selected event is examined using the 100 preceding stored flow fields together with the event snapshot. 
The time corresponding to snapshot $m$ is
\begin{equation}
    t_m
    =
    t_e+(m-101)\Delta t_{\mathrm{snap}},
    \qquad
    m=1,\ldots,101,
    \label{eqn:high_Re_snapshot_time}
\end{equation}
where $\Delta t_{\mathrm{snap}}$ is the interval between stored flow
fields and $m=101$ denotes the selected drag maximum. 
This interval allows the vortices contributing to the final trailing-edge interaction to be followed from approximately $x/c=0.5$ through the corresponding excursion in $C_D$.

\begin{figure}
\centering
\begin{minipage}{\linewidth}
    \begin{minipage}[c]{\linewidth}
        \centering 
        \includegraphics[trim=0.48cm 0.82cm 15cm 0.5cm, angle=90, clip, width=\linewidth]        {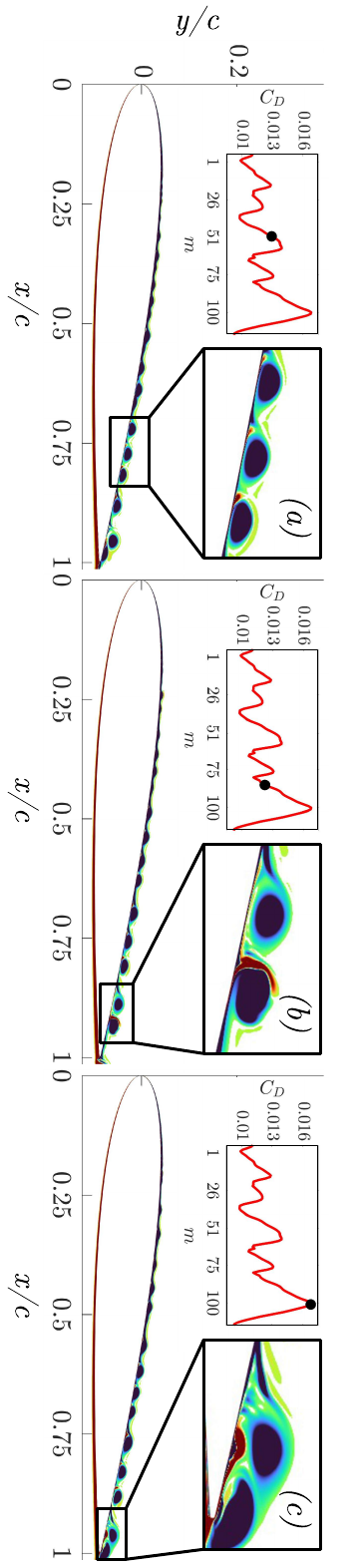}        
        \vspace{1em}
         (a)-(c): $EE_{T}-1$
    \end{minipage}
\end{minipage}
\vspace{0.5em}

\begin{minipage}{\linewidth}
    \begin{minipage}[c]{\linewidth}
        \centering
        \includegraphics[trim=0.48cm 0.82cm 15cm 0.5cm, angle=90, clip, width=\linewidth]        {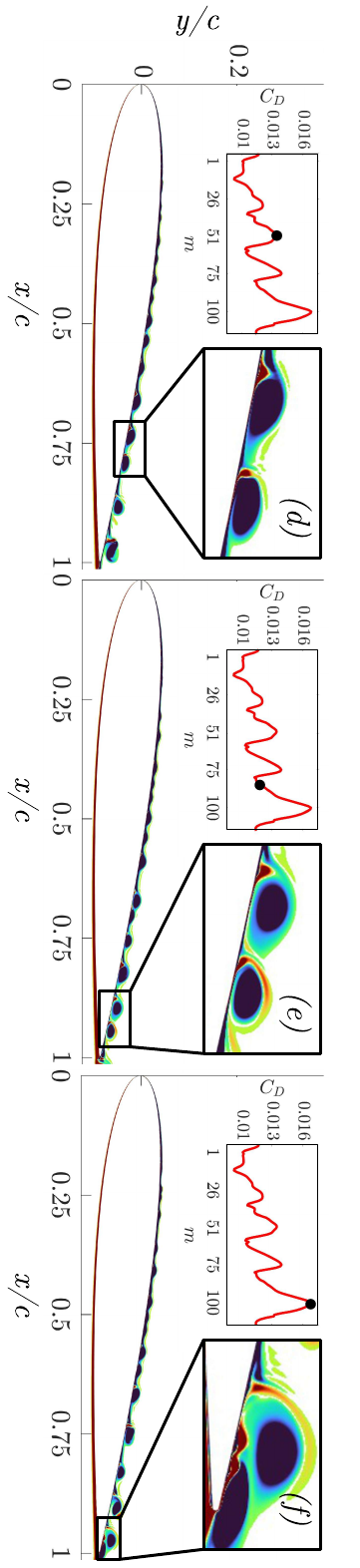}        
        \vspace{1em}
         (d)-(f): $EE_{T}-2$
    \end{minipage}
\end{minipage}
\vspace{0.5em}

\begin{minipage}{\linewidth}
    \begin{minipage}[c]{\linewidth}
        \centering
        \includegraphics[trim=0.41cm 0.82cm 15cm 0.5cm, angle=90, clip, width=\linewidth]        {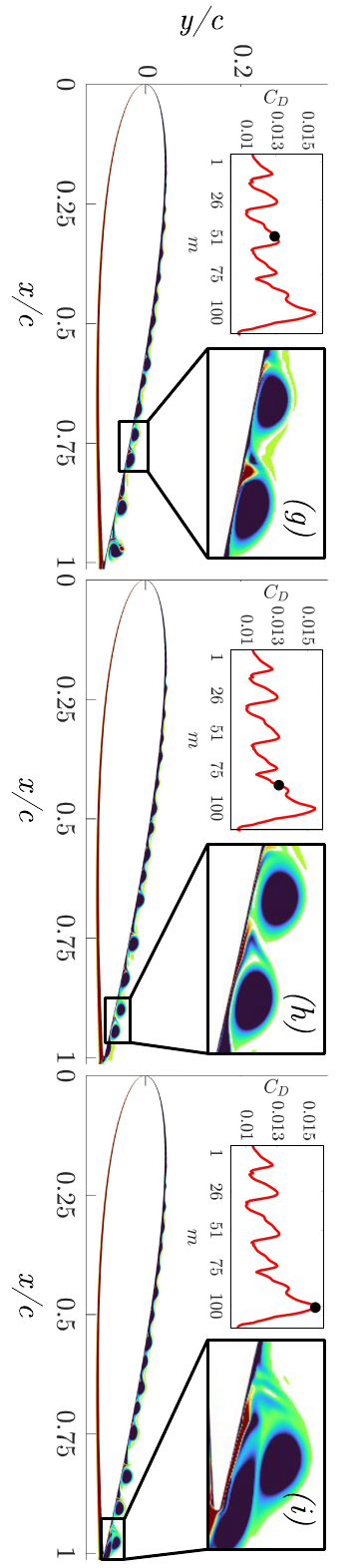}        
        \vspace{1em}
         (g)-(i): $EE_{T}-3$
    \end{minipage}
\end{minipage}
\vspace{0.5em}


\begin{minipage}{\linewidth}
    \begin{minipage}[c]{\linewidth}
        \centering 
        \includegraphics[trim=92cm 6cm 13cm 2cm, clip, angle = 90, width=\linewidth]        {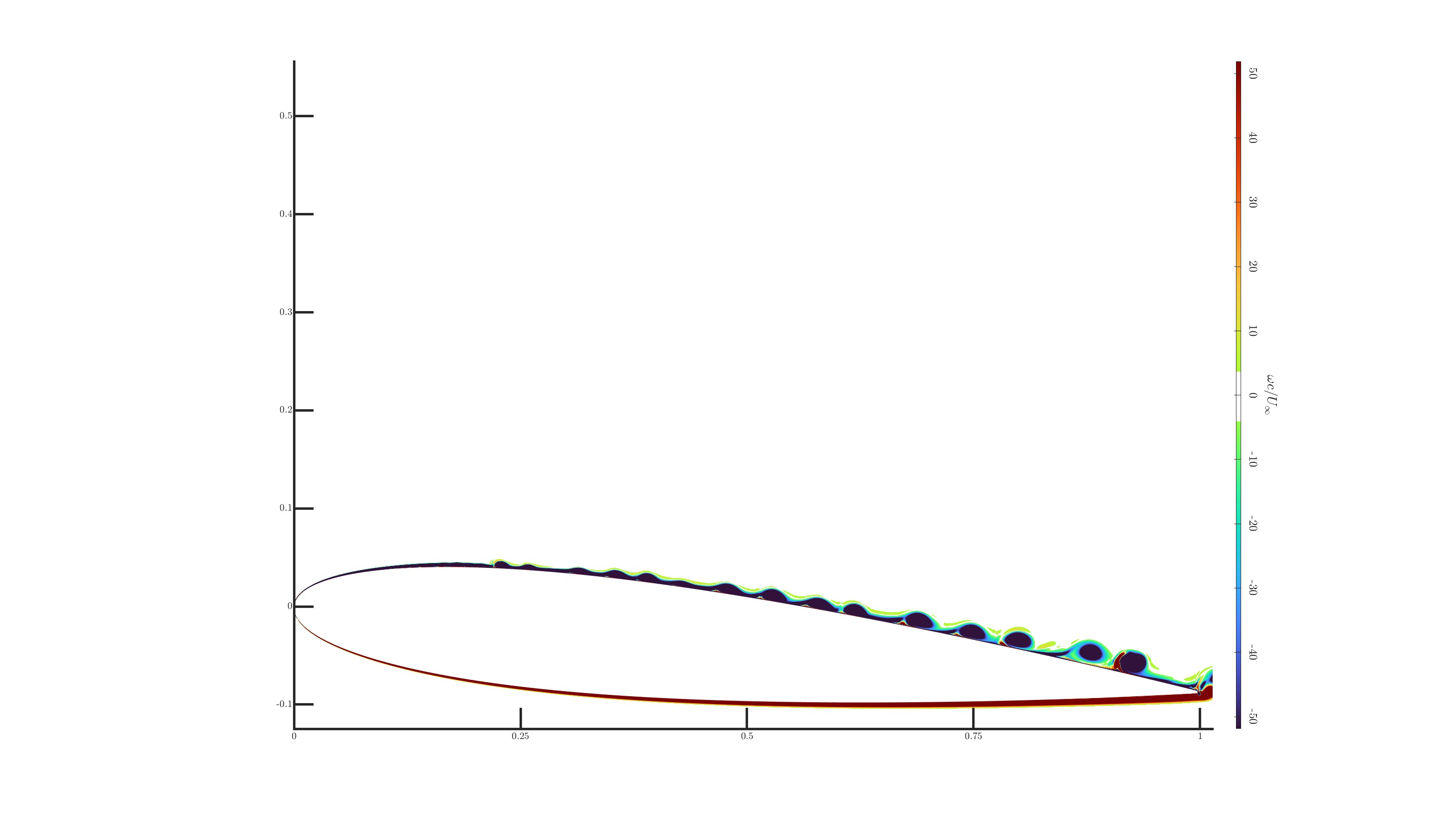}  
    \end{minipage}
\end{minipage}
\caption{Representative pre-event vortex evolution for the three
largest drag excursions in case $C_2$ at $Re=500{,}000$:
(a--c) $EE_T$--1, (d--f) $EE_T$--2, and (g--i) $EE_T$--3. The final column
corresponds to the selected drag maximum at $m=101$. Insets show the
corresponding $C_D$ histories.}
\label{fig:High_Re_vort_snap}
\end{figure}

Representative instantaneous-vorticity sequences are shown in
Fig.~\ref{fig:High_Re_vort_snap} and the corresponding complete pre-event sequences for representative cases of $EE_T$ and $NE$ are provided in supplementary movie~2.
Despite the rounded, finite-thickness trailing edge used in $C_2$, the same trailing-edge vortex-merging sequence is recovered and remains associated with pronounced excursions in $C_D$.
The selected extreme events contain at least three vortices that reach the aft portion of the airfoil with sufficiently small streamwise spacing to undergo a strong collective interaction. Several of the extreme events involve four vortices. 
The selected nominal states, in contrast, contain either a single vortex passing the trailing edge or an interaction involving no more than two vortices. 
This distinctiveness in the observed dynamics of vortices appearing near the trailing edge during the time of different events ($t_e$) is more clearly evident in the supplementary movie~2. Thus, although the individual vortices are smaller at the higher $Re$, their collective interaction can still generate an appreciable aerodynamic response when a sufficiently large and compact cluster appears near the trailing edge.

The resulting vortex cluster produces a localized pressure minimum over
the aft suction surface, similar to that observed during the extreme
events in $C_1$. The magnitude of the drag excursion is therefore not
determined solely by the strength of an individual vortex. It also
depends on how several vortices are organized when they approach the
trailing edge and on whether their pressure footprints combine to
produce a strong localized suction region.

The observations from $C_2$ extend the event pathway identified at
$Re=50{,}000$. In $C_1$, temporally clustered secondary-vorticity
eruptions release two or three discrete PVs with sufficiently small
spacing to interact near the trailing edge. In $C_2$, the separated
shear layer generates a larger population of smaller structures, and an
extreme event occurs when several of these structures remain closely
spaced and arrive near the trailing edge as a compact group. Thus,
although the detailed release process is not examined for $C_2$, both
cases exhibit the same downstream progression: a compact group of
vortices develops over the aft portion of the airfoil, undergoes a
strong collective interaction near the trailing edge, and produces a
localized suction region associated with the extreme drag excursion.

\subsection{Anticipated Effects of Three-Dimensionality}
\label{sec:spanwise_control_implications}
The present simulations isolate the extreme-event pathway in its spanwise-coherent limit. 
In this limit, the vortices and their associated surface-pressure signatures are uniform across the span,
providing a reference for the strongest span-integrated response that can arise from the near-trailing-edge interaction. 
The two-dimensional configuration therefore does more than simplify the flow: it exposes the temporal sequence linking vortex formation, interaction, and the resulting drag excursion without the additional modulation introduced
by spanwise dynamics.

In a three-dimensional flow, spanwise deformation, vortex stretching, and secondary instabilities may modify the strength, timing, and coherence of the interacting vortices. 
These processes do not necessarily eliminate the near-trailing-edge merger identified here.
Rather, they determine the spanwise extent over which the interaction remains organized. 
For instance, in the three-dimensional LES of
\citet{ricciardi2022transition}, vortices shed from the suction-side separation bubble exhibited two competing outcomes. 
Successful pairing produced a coherent structure that persisted toward the trailing edge, whereas unsuccessful pairing led to vortex bursting, transition, and a
less-coherent turbulent packet. 
Three-dimensionality therefore introduces pathways that can either preserve or disrupt the organized vortex interaction.

The mechanism identified here consequently remains available in a three-dimensional flow when sufficiently strong and closely spaced vortices interact near the trailing edge and remain correlated over an appreciable spanwise extent. 
A merger that remains coherent over a substantial portion of the span could produce a pronounced span-integrated pressure and drag excursion similar to that observed in
the present simulations. 
Even when the interaction is localized, it may still generate a strong sectional aerodynamic response.

The spanwise coherence length of the interacting vortex system thus emerges as an additional parameter governing extreme-event severity in three-dimensional flow. 
Future force-resolved, three-dimensional calculations should determine whether the near-trailing-edge interaction remains locally correlated, how rapidly the vortices lose spanwise coherence, and what spanwise extent is required to produce an appreciable integrated force excursion.

\section{Conclusion}
\label{sec: Conclusions}

The present results show that extreme drag excursions in transitional
airfoil flow are associated with intermittent reorganization of the
vortices generated by the separated shear layer. 
While Kelvin--Helmholtz roll-up, vortex shedding, and small-scale interactions occur routinely, the extreme events are distinguished by the arrival of a compact group of discrete vortices over the aft portion of the airfoil. 
Their strong collective interaction produces a concentrated near-trailing-edge pressure footprint and the corresponding sharp
increase in $C_D$.

The origin of this sequence lies in the release of primary vortices from the feeding shear layer. 
While a developing primary vortex remains connected to the shear layer, it continues to accumulate circulation and induces an opposite-signed secondary-vorticity layer at the wall.
Eruption of this secondary vorticity interrupts the connection to the
feeding shear layer and releases the primary vortex downstream. 
This release mechanism occurs during both nominal and extreme shedding. The distinguishing feature of the extreme events is the occurrence of
several such eruptions within a short interval. This clustered vortex
release establishes the small initial streamwise spacing that enables
the subsequent collective interaction.

The downstream evolution exhibits several features of classical
co-rotating vortex dynamics. 
Closely spaced vortices deform one another, develop differential convection, and may ultimately coalesce. 
The relative convection can develop through different temporal pathways, and a fixed streamwise ordering of vortex circulation is not required for strong interaction. 
The interactions themselves range from rapid core coalescence to prolonged stretching and filamentation. 
Thus, complete merger is not a prerequisite for the extreme response; a sufficiently compact vortex system interacting near the trailing edge can produce the concentrated pressure loading associated with the drag excursion.

The higher-$Re$ calculation provides an important extension of this
interpretation. As expected, the separated shear layer at
$Re=500{,}000$ produces a larger population of smaller vortical
structures. More notably, these smaller structures can still remain
closely spaced, undergo successive interactions, and produce pronounced
drag excursions near the trailing edge. The shorter $C_2$ record does
not permit a statistically converged comparison of event magnitude or
occurrence rate between the two Reynolds numbers. Nevertheless, the
results show that the identified downstream pathway persists after an
order-of-magnitude increase in $Re$ and with a rounded trailing edge with a finite thickness.

The present calculations describe the spanwise-coherent limit of this
process. In three-dimensional flow, spanwise deformation, vortex
stretching, and localized breakdown may weaken or localize the
interaction before it reaches the trailing edge. Conversely, where the
vortices retain sufficient spanwise coherence, the same local pathway
remains available. The spanwise extent of the interacting vortex system
is therefore expected to influence whether a strong sectional response
also produces a pronounced integrated force excursion.

The identified mechanism also suggests two opportunities for modifying the event sequence. 
Upstream, actuation that alters the local pressure-gradient history or the transport of near-wall secondary vorticity could modify the timing of primary-vortex release and disrupt the clustered-release process. 
Farther downstream, control that changes the relative convection or spacing of the released vortices could weaken their collective interaction near the trailing edge. 
These possibilities remain to be tested, but they suggest that extreme loading may be mitigated without suppressing Kelvin--Helmholtz roll-up itself, by instead targeting the organization of the vortices that emerge from it.


\section*{Acknowledgments}

This work was supported by the Office of Naval Research (ONR) under Grant No.~N00014-26-1-2288, with Dr.~Leighton Myers serving as the Program Officer. 
The views and conclusions expressed herein are those of the
authors and do not necessarily reflect the official policies or positions of ONR or the U.S. Government. 
The authors also acknowledge the computational resources provided by the High Performance Computing Center at Oklahoma State University, which is supported in part by the National Science Foundation under Grant No.~OAC-1531128.

\bibliography{JOURNAL_VERSION/Reference}
\end{document}